\documentclass[prd,superscriptaddress,twocolumn]{revtex4}

\usepackage[utf8]{inputenc}
\usepackage{amsmath}
\usepackage{graphicx}
\usepackage[colorinlistoftodos]{todonotes}
\usepackage{braket}
\usepackage{amssymb}
\usepackage{minitoc}
\usepackage{refstyle}
\usepackage{tensor}
\usepackage{bm}
\usepackage[caption=false]{subfig}

\usepackage{hyperref}
\hypersetup{
    hidelinks = true
}

\AtBeginDocument{\providecommand\eqref[1]{\ref{eq:#1}}}
\AtBeginDocument{\providecommand\secref[1]{\ref{sec:#1}}}
\AtBeginDocument{}
\AtBeginDocument{\providecommand\figref[1]{\ref{fig:#1}}}
\AtBeginDocument{\providecommand\tabref[1]{\ref{tab:#1}}}

\global\long\def\colmagenta#1{#1}

{\tiny{}}\global\long\def\braket#1{\Braket{#1}}
{\tiny \par}

\global\long\def\braOket#1#2#3{\left\langle #1\middle|#2\middle|#3\right\rangle }

{\tiny{}}\global\long\def\oq{\omega_{\rm q}}
{\tiny{}}\global\long\def\dxzp{\Delta x_{\rm zp}}
{\tiny{}}\global\long\def\osn{\omega_{\rm SN}}
{\tiny{}}\global\long\def\ocm{\omega_{\mathrm{cm}}}
{\tiny{}}
{\tiny{}}\global\long\def\ocavity{\omega_{\rm c}}
{\tiny{}}\global\long\def\aopt{\alpha}
{\tiny{}}\global\long\def\Iin{I_{{\rm in}}}

{\tiny{}}
{\tiny{}}\global\long\def\Gth{\Gamma}
{\tiny{}}\global\long\def\bMStr{\beta}

{\tiny{}}
{\tiny{}}

{\tiny{}}\global\long\def\Temp{T_{0}}

{\tiny{}}\global\long\def\Ftotth#1{\hat{F}_{{\rm th}}\left(#1\right)}
{\tiny{}}\global\long\def\fth#1{f_{{\rm cl}}\left(#1\right)}
{\tiny{}}\global\long\def\fzp#1{\hat{f}_{{\rm zp}}\left(#1\right)}
{\tiny{}}\global\long\def\fthF#1{f_{{\rm cl}}\left(#1\right)}
{\tiny{}}\global\long\def\Ftotthi{\hat{F}_{{\rm th}}}

{\tiny{}}
{\tiny{}}\global\long\def\dxcm{\Delta x_{{\rm cm}}}

{\tiny{}}\global\long\def\estimator{Y}

{\tiny{}}\global\long\def\SRQ#1{S_{\rm RQ}\left(#1\right)}
{\tiny{}}\global\long\def\SthQ#1{S_{f_{th},f_{th}}^{\rm qu}\left(#1\right)}

{\tiny{}}
{\tiny{}}
{\tiny{}}\global\long\def\Sxth#1{S_{x,x}^{\rm cl}\left(#1\right)}
{\tiny{}}\global\long\def\SbQ#1{S_{b_{2},b_{2}}^{\left({\rm QM}\right)}\left(#1\right)}
{\tiny{}}\global\long\def\SpreTot#1{ S_{b_{2},b_{2}}^{\left({\rm pre}\right)} \left(#1\right)}
{\tiny{}}\global\long\def\SpostTot#1{S_{b_{2},b_{2}}^{\left({\rm post}\right)}\left(#1\right)}

{\tiny{}}\global\long\def\SpostTot#1{S_{b_{2},b_{2}}^{\left({\rm post}\right)}\left(#1\right)}

{\tiny{}}\global\long\def\tmin{\tau_{{\rm min}}}

{\tiny{}}\global\long\def\postSig#1{D\left(#1\right)}

{\tiny{}}\global\long\def\Gc{G_{c}}

\newcommand{\specialcell}[2][c]{%
  \begin{tabular}[#1]{@{}c@{}}#2\end{tabular}}

\begin{document}

\title{Measurable signatures of quantum mechanics in a classical spacetime}

\makeatletter
\let\toc@pre\relax
\let\toc@post\relax
\makeatother 
\author{Bassam Helou}
\affiliation{Institute of Quantum Information and Matter, and Burke Institute of Theoretical Physics, M/C 350-17, California Institute of Technology, Pasadena, CA 91125, USA}
\author{Jun Luo}
\affiliation{Sun Yat-Sen University, Guangzhou, Guangdong, China}
\author{Hsien-Chi Yeh}
\affiliation{Sun Yat-Sen University, Guangzhou, Guangdong, China}
\author{Cheng-gang Shao}
\affiliation{Sun Yat-Sen University, Guangzhou, Guangdong, China}
\author{B J. J. Slagmolen}
\affiliation{Australian National University, Canberra, ACT, Australia}
\author{David E.\ McClelland}
\affiliation{Australian National University, Canberra, ACT, Australia}
\author{Yanbei Chen}
\affiliation{Institute of Quantum Information and Matter, and Burke Institute of Theoretical Physics, M/C 350-17, California Institute of Technology, Pasadena, CA 91125, USA}

\begin{abstract}
We propose an optomechanics experiment that can search for signatures of a fundamentally classical theory of gravity and in particular of the many-body Schroedinger-Newton (SN) equation, which governs the evolution of a crystal under a self-gravitational field. The SN equation predicts that the dynamics of a macroscopic mechanical oscillator's center of mass wavefunction differ from the predictions of standard quantum mechanics \cite{HuanPaper}. This difference is largest for low-frequency oscillators, and for materials, such as Tungsten or Osmium, with small quantum fluctuations of the constituent atoms around their lattice equilibrium sites. Light probes the motion of these oscillators and is eventually measured in order to extract valuable information on the pendulum's dynamics. Due to the non-linearity contained in the SN equation, we analyze the fluctuations of measurement results differently than standard quantum mechanics. We revisit how to model a thermal bath, and the wavefunction collapse postulate, resulting in two prescriptions for analyzing the quantum measurement of the light. We demonstrate that both predict features, in the outgoing light's phase fluctuations' spectrum, which are separate from classical thermal fluctuations and quantum shot noise, and which can be clearly resolved with state of the art technology.
\end{abstract}


\maketitle

\section{Introduction}

  
Advancements in quantum optomechanics has allowed the preparation, manipulation and characterization of the quantum states of macroscopic objects ~\cite{cavOptomechanics,meystre2013short,yanbeiReview}.  Experimentalists now have the technological capability to test whether gravity could modify quantum mechanics.  One option is to consider whether gravity can lead to decoherence,  as conjectured by Diosi and Penrose~\cite{penrose1996gravity,diosi1988quantum}, where the gravitational field around a quantum mechanical system can be modeled as being continuously monitored.   A related proposal is the Continuous Spontaneous Localization (CSL) model, which postulates that a different mass-density sourced field is being continuously monitored~\cite{bassi2013models}.   In both cases, gravity could be considered as having a ``classical component'', in the sense that transferring quantum information through gravity could be impeded, or even forbidden~\cite{kafri2014classical}. Another option, proposed by P.C.E. Stamp, adds gravitational correlations between quantum trajectories \cite{Stamp:2015vxa}.

In this paper, we consider a different, and more dramatic modification,
where the gravitational interaction is kept classical. Specifically,
the space-time geometry is sourced by the quantum expectation
value of the stress energy tensor~\cite{Rosenfeld1963353,MollerGravitation,Carlip:2008zf}:
\begin{equation}
G_{\mu\nu}=8\pi\left\langle \Phi|\hat{T}_{\mu\nu}|\Phi\right\rangle ,\label{eq:GT}
\end{equation}
with $G=c=1$, and where $G_{\mu\nu}$ is the Einstein tensor of a
(3+1)-dimensional classical spacetime.  $\hat{T}_{\mu\nu}$ is the
operator representing the energy-stress tensor, and $\left|\Phi\right\rangle $
is the wave function of all (quantum) matter and fields that evolve within this classical
spacetime. Such a theory arises either when researchers  considered gravity to be \textit{fundamentally classical}, or when
they ignored quantum fluctuations in the stress energy tensor, $T_{\mu\nu}$,
in order to approximately solve problems involving quantum gravity.
The latter case is referred to as \textit{semiclassical gravity}~\cite{Hu:2008rga},
in anticipation that this \textit{approximation} will break down if
the stress-energy tensor exhibits substantial quantum fluctuations.
In this article, we propose an optomechanics experiment that would test \eqref[name=Eq.~]{GT}. Other experiments have been proposed \cite{Grossardt:2015moa,craigSN}, but they do not address the difficulties discussed below.
 

Classical gravity, as described by \eqref[name=Eq.~]{GT},  suffers from a dramatic conceptual drawback rooted in the statistical interpretation of wavefunctions. In order for the Bianchi identity to hold
on the left-hand side of \eqref[name=Eq.~]{GT}, the right-hand side must be divergence free, but that would be violated if we reduced the quantum state. In light of this argument, one can go back to an interpretation
of quantum mechanics where the wavefunction does not reduce. At this moment, the predominant interpretation of quantum mechanics that does not have wave-function reduction is the relative-state, or ``many-world'' interpretation, in which all possible measurement outcomes, including macroscopically distinguishable ones, exist in the wavefunction of the universe. Taking an expectation over that wavefunction leads to a serious violation of common sense, as was demonstrated by Page and Geilker~\cite{page1981indirect}.

Another major difficulty is \textit{superluminal} communication, which follows from the nonlinearities implied by \eqref[name=Eq.~]{GT} (refer to \secref{SN} for explicit examples of nonlinear Schroedinger equations).  
Superluminal communication is a \textit{general} symptom of wavefunction collapse in nonlinear quantum mechanics \footnote{ We note that the issue of superluminal communication could be resolved by adding a stochastic extension to the theory of classical gravity, as was proposed by Nimmrichter~\cite{Nimmrichter:stochasticExtension}. However, although the theory removes the nonlinearity at the ensemble level, it also eliminates the signature of the nonlinearity in the noise spectrum. }.
Entangled and identically prepared states, distributed to two spatially separated parties $A$ and $B$, and then followed by projections at $B$ and a period of nonlinear evolution at $A$, can be used to transfer signals superluminally~\cite{polchinski1991weinberg,BassiSuperluminal,gisinNonlinear,SimonNoSignaling}.

In this paper, we do not solve the above conceptual obstacles. Instead,we highlight an \textit{even more serious issue} of nonlinear quantum mechanics: its dependence on the formulation of quantum mechanics. 
Motivated by the \textit{time-symmetric formulation} of quantum mechanics~\cite{reznik1995time},
we show that there are multiple prescriptions of assigning the probability
of a measurement outcome, that are equivalent in standard quantum
mechanics, but become distinct in nonlinear quantum mechanics.
It is our hope that at least one such formulation will not lead to
superluminal signaling. We defer the search for such a formulation to future work, and in this paper, we simply choose two prescriptions, and show that they give different experimental signatures in torsional pendulum experiments. These signatures hopefully scope out the type of behavior classical gravity would lead to if a non superluminal-signaling theory indeed exists.


This paper is organized as follows.  In section \ref{sec:SN}, we review the non-relativistic limit of \eqref[name=Eq.~]{GT}, called the \textit{Schroedinger-Newton theory}, as applied to optomechanical setups, and without including quantum measurements. We determine that the signature of the Schroedinger-Newton theory in the free dynamics of the test mass is largest
for low frequency oscillators such as torsion pendulums, and for materials, such as Tungsten and Osmium, with atoms tightly bound around their respective lattice sites.
In  section \ref{sec:Modeling-the-bath}, we remind the reader that in nonlinear quantum mechanics the density matrix formalism cannot be used to describe thermal fluctuations. As a result, we propose a particular ensemble of pure states to describe the thermal bath's state. In section \ref{sec:Measurements-in-NLQM}, we discuss two strategies, which we term pre-selection and post-selection, for assigning a statistical interpretation to the wavefunction in the Schroedinger-Newton theory. In section \ref{sec:signatures}, we obtain the signatures of the pre- and post-selection prescriptions in torsional pendulum experiments.  In section \ref{sec:Feasibility-analysis}, we show that is feasible to measure these signatures in state of the art experiments. Finally, we summarize our main conclusions in section \ref{sec:conclusions}.

\section{Free dynamics of an optomechanical setup under the Schroedinger-Newton theory}
\label{sec:SN}

In this section, we discuss the Schroedinger-Newton theory applied to optomechanical setups without quantum measurement. We first review the signature of the theory in the free dynamics of an oscillator, and discuss associated design considerations. We then  develop an effective Heisenberg picture, which we refer to as a \textit{state dependent Heisenberg picture}, where only operators evolve in time. However, unlike the Heisenberg picture, the equations of motion depend on the boundary quantum state  of the system that is being analyzed. Finally we present the equations of motion of our proposed optomechanical setup.

\subsection{\label{comSNEq}The center-of-mass Schroedinger-Newton equation}

The Schrödinger-Newton theory follows from taking the non-relativistic limit of \eqref[name=Eq.~]{GT}.  The expectation value in  this equation gives rise to a nonlinearity. In particular, a single non-relativistic particle's wavefunction, $\chi\left(\vec{r}\right)$, evolves as
\begin{equation}
i\hbar\partial_{t}\chi\left(\vec{r},t\right)=\left[-\frac{\hbar^{2}}{2m}\nabla^{2}+V\left(\vec{r}\right)+U\left(t,\vec{r}\right)\right]\chi\left(\vec{r},t\right),
\end{equation}
where $V\left(\vec{r}\right)$ is the non-gravitational potential energy at $\vec{r}$ and  $U\left(t,\vec{r}\right)$ is the Newtonian self-gravitational potential and is sourced by $\chi\left(\vec{r}\right)$:
\begin{equation}
\nabla^{2}U\left(t,x\right)=4\pi G m\left|\chi\left(t,x\right)\right|^{2} .
\end{equation}

A many-body system's center of mass Hamiltonian also admits a simple description, which was analyzed in \cite{HuanPaper}.  If an object has its center of mass' displacement fluctuations much smaller than fluctuations of the internal motions of its constituent atoms, then its center of mass, with quantum state  $|\psi\rangle$, observes
\begin{equation}
i\hbar\frac{d\left|\psi\right\rangle}{dt} =\left[\hat H_{\rm NG}+\frac{1}{2}M\osn^{2}\left(\hat{x}-\left\langle \psi|\hat{x}|\psi\right\rangle \right)^{2}\right]\left|\psi\right\rangle \label{eq:evolutionCMsnintro}
\end{equation}
where $M$ is the mass of the object, $\hat H_{\rm NG}$ is the non-gravitational part of the Hamiltonian, $\hat x$ is the center of mass position operator, and $\osn$ is a frequency scale that is determined by the matter distribution of the object.  For materials with single atoms sitting close to lattice sites, we have 
\begin{equation}
{\osn}\equiv\sqrt{\frac{Gm}{6\sqrt{\pi}\dxzp^{3}}}
\end{equation}
where $m$ is the mass of the atom, and $\dxzp$ is the standard deviation of the  crystal's constituent atoms' displacement from their equilibrium position along each spatial direction due to quantum fluctuations. 

Note that the presented formula for $\osn$ is larger than the expression for $\osn$ presented in \cite{HuanPaper} by a factor of $\sqrt{2}$. As explained in \cite{Grossardt:SNfactor12}, the many body non-linear gravitational interaction term presented in Eq. (3) of \cite{HuanPaper} should not contain a factor of 1/2, which is usually introduced to prevent overcounting. The SN interaction term between one particle and another is not symmetric under exchange of both them. For example, consider two  (1-dimensional) identical particles of mass $m$. The interaction term describing  the gravitational attraction of the first particle, with position operator $\hat{x}_{1}$, to the second is given by
$$-Gm^{2}\int dx_1\;dx_2\frac{\left|\psi\left(x_1,x_2\right)\right|^2}{\left|\hat{x}_{1}-x_2\right|},$$
which is not symmetric under the exchange of the indices 1 and 2. Moreover, in Appendix \ref{sec:consEnergy}, we show that the expectation value of the total Hamiltonian is not conserved. Instead, 
\begin{equation}
E=\left\langle \hat{H}_{{\rm NG}}+\hat{V}_{{\rm SN}}/2\right\rangle 
\end{equation}
is conserved, where $\hat{V}_{{\rm SN}}$ is the SN gravitational potential term. As a result, we take $E$, which contains the factor of 1/2 present in expressions of the classical many-body gravitational energy, to be the average energy. 

 If the test mass is in an external harmonic potential, \eqref[name=Eq.~]{evolutionCMsnintro} becomes
 \begin{align}
i\hbar\frac{d\left|\psi\right\rangle}{dt} =&\bigg[\frac{\hat{p}^{2}}{2M}+\frac{1}{2}M\ocm^{2}\hat{x}^{2}\nonumber\\
&+\frac{1}{2}M\osn^{2}\left(\hat{x}-\left\langle \psi|\hat{x}|\psi\right\rangle \right)^{2}\bigg]\left|\psi\right\rangle
\label{eq:evolutionCMsn}
\end{align}
where $\hat{p}$ is the center of mass momentum operator, and $\colmagenta{\ocm}$ is the resonant frequency of the crystal's motion in the absence of gravity. 

\eqref[name=Eq.~]{evolutionCMsn} predicts distinct dynamics from linear quantum mechanics. Assuming a Gaussian initial state, Yang \emph{et al.} show that the signature of \eqref[name=Eq.~]{evolutionCMsn} appears in the rotation frequency
\begin{equation}
\colmagenta{\oq}\equiv\sqrt{\omega_{\rm cm}^{2}+\omega_{\rm SN}^{2}}\label{eq:defQfreq}
\end{equation}
of the mechanical oscillator’s quantum uncertainty ellipse in phase space. We illustrate this behavior in \figref[name=Fig.~]{2freq}. 

As a consequence, the dynamics implied by the nonlinearity in \eqref[name=Eq.~]{evolutionCMsnintro} are most distinct from the predictions of standard quantum mechanics when $\oq - \ocm$ is as large
as possible. This is achieved by having a pendulum with as small of an oscillation eigenfrequency as possible, and made with a material with as high of a $\osn$ as possible. The former condition
leads us to propose the use of low-frequency torsional pendulums. To meet the latter condition, we notice that $\osn$ depends significantly on $\dxzp$, which can be inferred from the Debye-Waller factor, 
\begin{equation}
\colmagenta B=u^{2}/8\pi{}^{2}
\end{equation}
where $u$ is the rms displacement of an atom from its equilibrium
position~\cite{Peng:zh0008}. Specifically, thermal and intrinsic fluctuations contribute to $u$, \emph{i.e.} $u^{2}\gtrsim\sqrt{\dxzp^{2}+\Delta x_{\rm th}^{2}}$
with $\Delta x_{\rm th}$ representing the uncertainty in the internal
motion of atoms due to thermal fluctuations. 

In \Tabref{DW}, we present experimental data on some materials' Debye-Waller factor, and conclude that 
the pendulum should ideally be made with Tungsten (W), with $\osn^{{\rm W}}=2\pi\times 4.04\,{\rm mHz}$, or Osmium (the densest naturally occurring element) with a theoretically predicted $\osn^{\rm Os}$ of $2\pi\times 5.49\,{\rm mHz}$.
Other materials such as Platinum or Niobium, with $\osn^{{\rm Pt}}=2\pi\times 3.2\,{\rm mHz}$
and $\osn^{{\rm Nb}}=2\pi\times 1.56\,{\rm mHz}$ respectively,
could be suitable candidates.

\begin{figure}
\centering\includegraphics[scale=0.7]{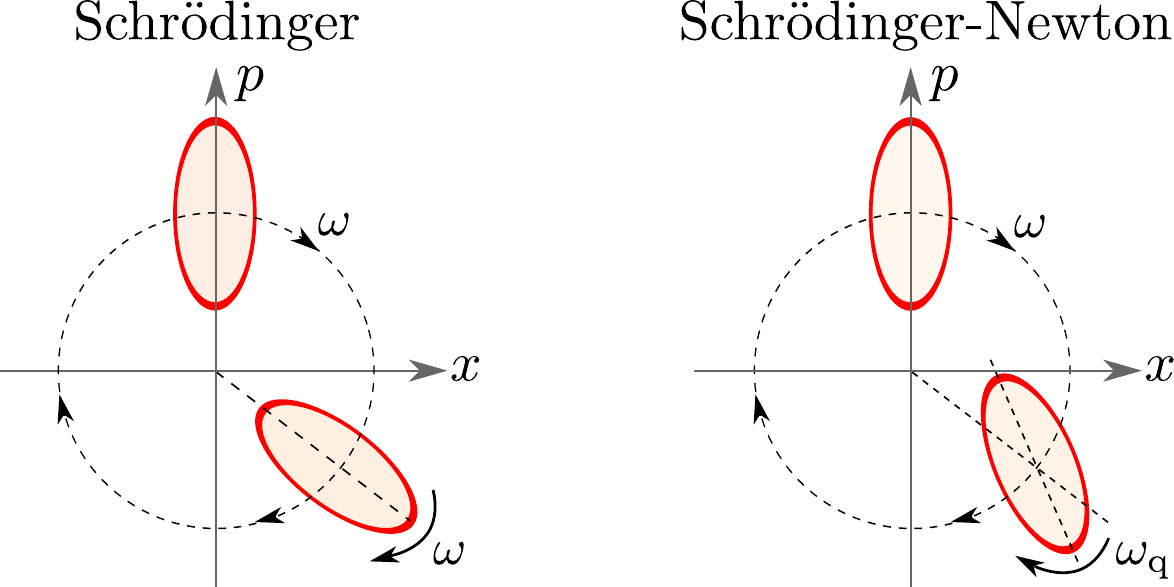}

\caption{\label{fig:2freq} Left Panel: according to standard
quantum mechanics, both the vector $(\langle \hat{x}\rangle,\langle \hat{p}\rangle)$
and the uncertainty ellipse of a Gaussian state for the center of mass of a macroscopic object rotate clockwise in phase space, at the same frequency $\omega=\omega_{{\rm CM}}$.
Right panel: according to  \eqref[name=Eq.]{evolutionCMsn},
$(\langle x\rangle,\langle p\rangle)$ still rotates at $\ocm$,
but the uncertainty ellipse rotates at $\oq\equiv\sqrt{{\ocm}^{2}+\osn^{2}}>\ocm$.
(Figure taken from \cite{HuanPaper}).}
\end{figure}

\begin{table}[t]
\begin{tabular}{cccc}
Element  & %
\begin{tabular}{c}
$\rho$\tabularnewline
$(10^{3}\,\mathrm{kg}/\mathrm{m}^{3})$\tabularnewline
\end{tabular} & %
\begin{tabular}{c}
$B^{2}$\tabularnewline
$(\AA^{2})$ \tabularnewline
\end{tabular} & %
\begin{tabular}{c}
$\omega_{{\rm SN}}$\tabularnewline
($10^{-2}\,s^{-1}$) \tabularnewline
\end{tabular}\tabularnewline
\hline 
\hline 
Silicon (Si)  & 2.33  & 0.1915  & $4.95$\tabularnewline
Iron (Fe) (BCC) & 7.87  & 0.12 & $9.90$\tabularnewline
Germanium (Ge) & 5.32 & 0.1341 & $10.39$\tabularnewline
Niobium (Nb)  & 8.57  & 0.1082  & $13.86$ \tabularnewline
Platinum (Pt)  & 21.45 & 0.0677  & $28.43$ \tabularnewline
Tungsten (W)  & 19.25  & 0.0478  & $35.92$\tabularnewline
Osmium{*} (Os) & 22.59  & 0.0323  & $48.79$ \tabularnewline
\end{tabular}\caption{Characteristic Schroedinger-Newton
angular frequency $\omega_{{\rm SN}}$ for several elemental crystals.
Density is approximated by values at room temperature, and the Debye-Waller
factor $B$ (at 1\,K) is provided by Ref.~\cite{Peng:zh0008}.
{*}: Note that Osmium's Debye-Waller factor is solely obtained from
theoretical calculations.\label{tab:DW}}
\end{table}

\subsection{State-dependent Heisenberg picture for nonlinear quantum mechanics}

In this section, we develop an effective Heisenberg picture for non-linear Hamiltonians similar to the Hamiltonian given by \eqref[name=Eq.]{evolutionCMsn}. We abandon the Schroedinger picture because the dynamics of a Gaussian optomechanical system are usually examined in the Heisenberg picture where the similarity to classical equations of motion is most apparent.

We are interested in non-linear Schroedinger equations of the form
\begin{align}
i\hbar \frac{d|\psi\rangle}{dt} &= \hat H(\zeta(t)) |\psi\rangle\,,
\label{eq:psiNLevo}
\\
\zeta(t) & = \langle \psi(t) |\hat Z |\psi(t)\rangle,
\end{align}
where the Hamiltonian $\hat{H}$ is a linear operator that depends on a parameter $\zeta$, which in turn depends on the quantum state that is being evolved. Note that the Schroedinger operator $\hat Z$ can depend explicitly on time, $\zeta$ can have multiple components, and the Hilbert space and canonical commutation relations are unaffected by the nonlinearities. 

\subsubsection{State-dependent Heisenberg Picture}

We now present the effective Heisenberg Picture. Let us identify the Heisenberg and Schroedinger pictures at the initial time $t=t_0$,
\begin{equation}
|\psi_H\rangle =|\psi(t_0)\rangle ,\quad  \hat x_H(t_0) =\hat x_S(t_0) ,\quad \hat p_H(t_0) =\hat p_S(t_0),
\label{eq:boundaryHeisS}
\end{equation}
where $|\psi_H\rangle$ is the quantum state $|\psi\rangle$ in the Heisenberg picture, and we have used the subscripts $S$ and $H$ to explicitly indicate whether an operator is in the Schroedinger or Heisenberg picture, respectively.
As we evolve (forward or backward) in time in the Heisenberg Picture, we fix $|\psi_S(t_0)\rangle$, but evolve $\hat x_H (t)$ according to
\begin{align}
\frac{d}{dt}\hat x_H (t) &= \frac{i}{\hbar}\left[\hat H_H(\zeta(t)),\hat x_H(t)\right]+\frac{\partial }{\partial t}\hat x_H(t) \,,\quad\\
\zeta(t) & = \langle \psi_H |\hat Z_H (t)|\psi_H\rangle.
\end{align}
A similar equation holds for $\hat p_H (t)$. We shall refer to such equations as {\it state-dependent Heisenberg equations of motion}.
Moreover, the Heisenberg picture of an arbitrary operator in the Schroedinger picture
\begin{equation}
\hat{O}_{S}=f\left(\hat{x}_{S},\hat{p}_{S},t\right),
\end{equation}
including the Hamiltonian $\hat H(\zeta(t))$, can be obtained from $\hat x_H (t)$ and  $\hat p_H (t)$ by: 
\begin{equation}
\hat{O}_{H}\left(t\right)=f\left(\hat{x}_{H}\left(t\right),\hat{p}_{H}\left(t\right),t\right).
\end{equation}

\subsubsection{Proof of the State-Dependent Heisenberg Picture}

The state-dependent Heisenberg picture is equivalent to the Schroedinger picture, if at any given time
\begin{equation}
\langle \psi_H |\hat O_H(t) |\psi_H \rangle  = \langle\psi_S(t) |\hat O_S(t) |\psi_S(t)\rangle .
\label{eq:effHeisToProve1}
\end{equation}
Before we present the proof, we motivate the existence of a Heisenberg picture with a simple argument. If we (momentarily) assume that the nonlinearity $\zeta\left(t\right)$
is known and solved for, then the non-linear Hamiltonian $\hat{H}\left(\zeta\left(t\right)\right)$
is mathematically equivalent to a linear Hamiltonian, 
\begin{equation}
\hat{H}^{L}\left(\zeta\left(t\right)\right)=\hat{H}\left(\zeta\left(t\right)\right),
\end{equation}
with a classical time-dependent drive $\zeta\left(t\right)$. Since there exists a Heisenberg picture associated with $\hat{H}^{L}\left(\zeta\left(t\right)\right)$, there exists one for the nonlinear Hamiltonian $\hat{H}\left(\zeta\left(t\right)\right)$.

We now remove the assumption that $\zeta\left(t\right)$ is known and consider linear Hamiltonians, $\hat{H}^{L}\left(\rho\left(t\right)\right)$, driven by general time-dependent classical drives $\lambda\left(t\right)$. To each $\hat{H}^{L}\left(\lambda\left(t\right)\right)$ is associated a different unitary operator $\hat{U}_{\lambda}\left(t\right)$ and so a different Heisenberg picture 
\begin{equation}
\hat{O}_{H}\left(\lambda, t\right) = \hat{U}_{\lambda}^{\dagger}\left(t\right)\hat{O}_{S}\hat{U}_{\lambda}\left(t\right).
\label{eq:HeisOh}
\end{equation}
Next, we choose $\lambda(t)$ in such a way that 
\begin{equation}
\langle \psi_H |\hat O_H(\lambda,t) |\psi_H \rangle  = \langle\psi_S(t) |\hat O_S(t) |\psi_S(t)\rangle .
\label{eq:effHeisToProve}
\end{equation}
is met. For the desired effective Heisenberg picture to be self-consistent, $\lambda\left(t\right)$ must be obtained by solving 
\begin{equation}
\lambda(t)=\left\langle \psi_{H}|\hat{Z}_{H}\left(\lambda, t\right)|\psi_{H}\right\rangle,
\label{eq:defRho}
\end{equation}
which, in general, is a non-linear equation in $\lambda$.
We will explicitly prove that this choice of $\lambda(t)$ satisfies \eqref[name=Eq.~]{effHeisToProve}. Note that we will present the proof in the case that the boundary wavefunction is forward time evolved. The proof for backwards time evolution is similar. 

We begin the proof by showing that $\lambda$ and $\zeta$ are equal at $t=t_{0}$, 
\begin{equation*}
\lambda \left(t_0 \right) =\left\langle \psi_{S}\left(t_{0}\right)|\hat{Z}_{S}|\psi_{S}\left(t_{0}\right)\right\rangle =\zeta\left(t_{0}\right)
\end{equation*}
because the Schrodinger and state-dependent Heisenberg pictures are, as indicated by \eqref[name = Eq.~]{boundaryHeisS}, identified at the initial time $t=t_0$. 

$\lambda$ and $\zeta$ can deviate at later times if the increments $\partial_t \lambda$ and $\partial_t \zeta$ are different. We use the nonlinear Schroedinger equation to obtain the latter increment:
\begin{eqnarray}
\partial_{t}\zeta\left(t\right) & = & \partial_{t}\left\langle \psi_{S}\left(t\right)|\hat{Z}_{S}|\psi_{S}\left(t\right)\right\rangle \\
 & = & \frac{i}{\hbar}\left\langle \psi_{S}\left(r\right)|\left[\hat{H}\left(\zeta\left(t\right)\right),\hat{Z}_{S}\right]|\psi_{S}\left(t\right)\right\rangle.
\end{eqnarray}
Note that the equation of motion for $\zeta\left(t\right)$ is particularly
simple to solve in the case of the quadratic Hamiltonian given by  \eqref[name=Eq.~]{evolutionCMsnintro}, because the non-linear part of $\hat{H}\left(\zeta\left(t\right)\right)$ commutes with $\hat{x}$.

On the other hand, by \eqref[name=Eq.~]{defRho},
\begin{eqnarray*}
\partial_{t}\lambda(t) &=&\frac{i}{\hbar}\left\langle \psi_{H}|\left[\hat{H}_{H}^{L}\left(\lambda\left(t\right)\right),\hat{Z}_{H}\left(\lambda, t\right)\right]|\psi_{H}\right\rangle 
\end{eqnarray*}
Making use of \eqref[name=Eq.~]{HeisOh}, and of 
\begin{equation}
\hat{H}^{L}\left(\lambda\left(t\right)\right)  =  \hat{U}_{\lambda}\left(t\right)\hat{H}_{H}^{L}\left(\lambda\left(t\right)\right)\hat{U}_{\lambda}^{\dagger}\left(t\right),
\end{equation}
we obtain 
\begin{eqnarray*}
\partial_{t} \lambda(t)  & = & \frac{i}{\hbar} \braOket{\hat{U}_{\lambda}\left(t\right)\psi_{H}}{\left[\hat{H}^{L}\left(\lambda\left(t\right)\right),\hat{Z}_{S}\right]}{\hat{U}_{\lambda}\left(t\right)\psi_{H}}.
\end{eqnarray*}
Furthermore, $\left|\hat{U}_{\lambda}\left(t\right)\psi_{H}\right\rangle$ evolves under 
\begin{equation}
i\hbar\frac{d\left|\hat{U}_{\lambda}\left(t\right)\psi_{H}\right\rangle }{dt}=\hat{H}\left(\lambda\left(r\right)\right)\left|\hat{U}_{\lambda}\left(t\right)\psi_{H}\right\rangle 
\end{equation}
Notice the similarity with \eqref[name=Eq.~]{psiNLevo}.

We have established that the differential equations governing the time evolution of $\lambda$ and $\left|\hat{U}_{\lambda}\left(t\right)\psi_{H}\right\rangle$, are of the same form as those governing the time evolution of $\zeta(t)$ and $\left|\psi_{S}\left(t\right)\right\rangle$. In addition, these equations have the same initial conditions. Therefore, $\lambda(t)=\zeta(t)$ for all times $t$. \eqref[name=Eq.~]{effHeisToProve} then easily follows because we've established that $\hat{H}\left(\zeta\left(t\right)\right)$ 
and 
\begin{equation*}
\hat{H}^{L}\left(\left\langle \psi_{H}|\hat{Z}_{H}\left(\lambda, t\right)|\psi_{H}\right\rangle \right)
\end{equation*}
are mathematically equivalent for all times $t$. 

\subsection{Optomechanics without measurements}

We propose to use laser light, enhanced by a Fabry-Perot cavity, to monitor the motion of the
test mass of a torsional pendulum, as shown in \figref[name=Fig.~]{The-proposed-setup}. 
We assume the light to be resonant with the cavity, and that the cavity has a much larger linewidth than $\oq$, the frequency of motion we are interested in. 

We will add the non-linear Schroedinger-Newton term from \eqref[name=Eq.~]{evolutionCMsn} to the usual optomechanics Hamiltonian, obtaining
\begin{eqnarray}
\hat H & = & \hat H_{\rm OM} + \frac{1}{2}M\omega_{\rm SN}^2 (\hat x -\braOket{\psi}{\hat{x}}{\psi})^2,
\label{eq:Hdef}
\end{eqnarray} 
where $\hat H_{\rm OM}$ is the standard optomechanics Hamiltonian for our system \cite{yanbeiReview}. We have ignored corrections due to light's gravity because we are operating in the Newtonian regime, where mass dominates the generation of the gravitational field.
$\hat H$ generates the following linearized state dependent Heisenberg equations (with the dynamics of the cavity field adiabatically eliminated, and the "H" subscript omitted):
\begin{align}
\partial_{t}\hat{x} & =\frac{\hat{p}}{M}\label{eq:xEOM}\\
\partial_{t}\hat{p} & =-M\ocm^{2}\hat{x}-M\omega_{\rm SN}^2(\hat x -  \braOket{\psi}{\hat{x}}{\psi}) +\alpha\hat{a}_{1}\label{eq:pEOM}\\
\hat{b}_{1} & =\hat{a}_{1} \\
\hat{b}_{2}&=\hat{a}_{2}+\frac{\alpha}{\hbar}\hat{x},
\end{align}
where $\hat a_{1,2}$ are the perturbed incoming quadrature fields around a large steady state, and similarly $\hat b_{1,2}$  are the perturbed outgoing field quadratures (refer to section 2 of \cite{yanbeiReview} for details). 
The quantity $\alpha$ characterizes the optomechanical coupling, and depends on the pumping power $\Iin$ and the input-mirror power transmissivity $T$ of the Fabry-Perot cavity:
\begin{equation}
\aopt^{2}=\frac{8\Iin}{T}\frac{\hbar\ocavity}{c^{2}}\frac{1}{T}.
\label{eq:alphaDef}
\end{equation}

Note that we have a {\it linear system} under {\it nonlinear quantum mechanics} because the Heisenberg equations are linear in the center of mass displacement and momentum operators, and in the optical field quadratures, including their expectation value on the system's quantum state.   

\begin{figure}
\centering\includegraphics[scale=0.8]{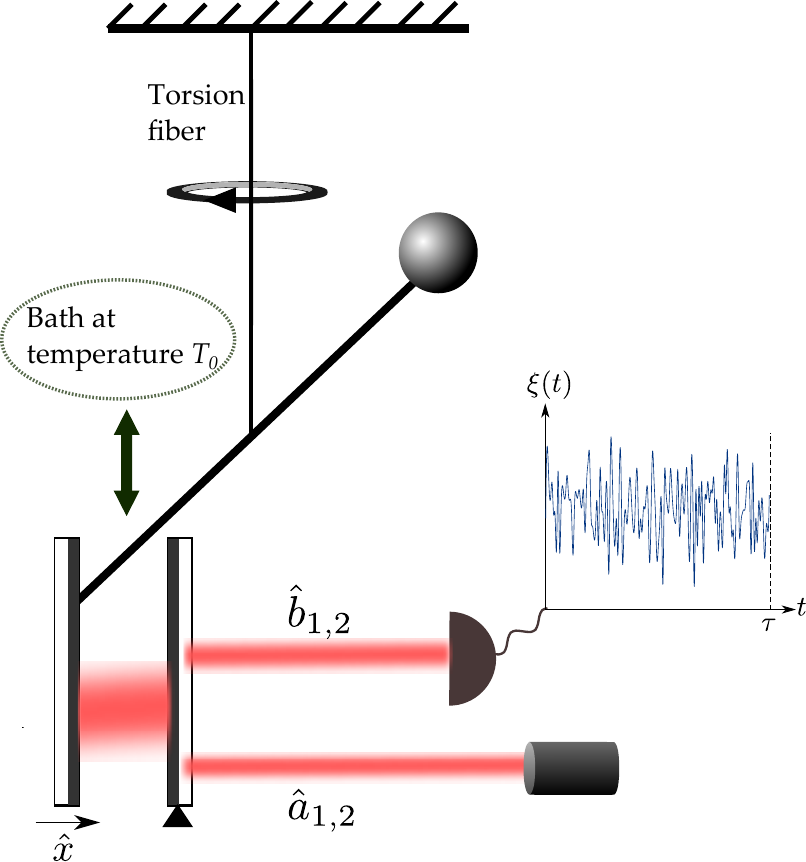}

\caption{\label{fig:The-proposed-setup}The proposed low-frequency optomechanical experiment. }
\end{figure}

\section{Nonlinear quantum optomechanics with classical noise}
{\label{sec:Modeling-the-bath}

To study realistic optomechanical systems, we must incorporate thermal fluctuations. In linear quantum mechanics, we usually do so by describing the state of the bath with a density operator. However, it is known that the density matrix formalism cannot be used in non-linear quantum mechanics~\cite{BassiSuperluminal}.

Our dynamical system is linear and is driven with light in a Gaussian state, so all system states are eventually Gaussian. Moreover, our system admits a state-dependent Heisenberg picture. Consequently, we can describe fluctuations with distribution functions of linear observables which are completely characterized by their first and second moments.
In nonlinear quantum mechanics, the challenge will be to distinguish between quantum uncertainty and the probability distribution of classical forces. The conversion of quantum uncertainty to probability distributions of measurement outcomes is a subtle issue in nonlinear quantum mechanics, and will be postponed until the next section. 

Once we have chosen a model for the bath, we will have to revisit the constraint, required for \eqref[name=Eq.~]{evolutionCMsn} to hold, that the center of mass displacement fluctuations are much smaller than $\dxzp$. Thermal fluctuations increase the uncertainty in the center of mass motion to the point that in realistic experiments, the total displacement of the test mass will be much larger than $\Delta x_{\rm zp}$.
Nonetheless, after separating classical and quantum uncertainties, we will show that Eq.~(\ref{eq:evolutionCMsn}) remains valid, as long as the quantum (and not total) uncertainty of the test mass is much smaller than $\Delta x_{\rm zp}$. 

Finally, we ignore the gravitational interactions in the thermal bath, as they are expected to be negligible.

\subsection{Abandoning the density matrix formalism in nonlinear quantum mechanics }

 In standard quantum mechanics, we use the density matrix formalism when a system is entangled with another system and/or when we lack information about a system's state.
The density matrix completely describes a system's quantum state. If two different ensembles of pure states, say $\left\{ \left|\psi_{i}\right\rangle \right\} $
and $\left\{ \left|\phi_{i}\right\rangle \right\} $ with corresponding
probability distributions $p_{\psi_{i}}$ and $p_{\phi_{i}}$, have the same density matrix 
\begin{equation}
\sum_{i}p_{\psi_{i}}\left|\psi_{i}\right\rangle \left\langle \psi_{i}\right|=\sum_{i}p_{\phi_{i}}\left|\phi_{i}\right\rangle \left\langle \phi_{i}\right|,
\end{equation}
then they cannot be distinguished by measurements. Furthermore, when either ensemble is  time-evolved, they will keep having the same density matrix. However, this statement is no longer true in non-linear quantum mechanics because the superposition principle is no longer valid.

Let us give an example of how our nonlinear Schroedinger equation, given by \eqref[name=Eq.~]{evolutionCMsn}, implies the breakdown of the density matrix formalism. 
Suppose Alice and Bob share a collection of entangled states, $\left|\Phi\right\rangle$, between Bob's test mass' center of mass degree of freedom  and Alice's spin 1/2 particle, with $\left|\Phi\right\rangle$ given by
\begin{eqnarray*}
\left|\Phi\right\rangle  & = & \frac{1}{\sqrt{2}}\left(\left|\uparrow\right\rangle \left|\psi_x\right\rangle +\left|\downarrow\right\rangle \left|\psi_{-x}\right\rangle \right)\\
 & = & \frac{1}
 {\sqrt{2}}\left(\left|\rightarrow\right\rangle \left|+\right\rangle +\left|\leftarrow\right\rangle \left|-\right\rangle \right)
\end{eqnarray*}
where
\begin{eqnarray}
\left|\rightarrow\right\rangle  & \equiv & \frac{\left|\uparrow\right\rangle +\left|\downarrow\right\rangle }{\sqrt{2}}\\
\left|\leftarrow\right\rangle  & \equiv & \frac{\left|\uparrow\right\rangle -\left|\downarrow\right\rangle }{\sqrt{2}},
\end{eqnarray}
and $\left|\psi_{\pm x}\right\rangle$ are localized states around $x$ and $-x$:
\begin{equation}
\left|\psi_{\pm x}\right\rangle =\frac{1}{\sqrt{\sigma\sqrt{\pi}}}\int\exp\left(-\frac{\left(y\mp x\right)^{2}}{2\sigma^2}\right)\left|y\right\rangle dy.
\end{equation}
We choose $\sigma \ll x$ so that $\left\langle \psi_{x}|\psi_{-x}\right\rangle \approx0$.
Moreover,
\begin{eqnarray}
\left| \pm \right\rangle  & \equiv & \frac{1}{\sqrt{2}}\left(\left|\psi_{x}\right\rangle \pm \left|\psi_{-x}\right\rangle \right).
\end{eqnarray}
Next, suppose that Alice measures her spins along the $\left\{ \left|\uparrow\right\rangle ,\left|\downarrow\right\rangle \right\} $ basis, then Bob will be left with the following mixture of states:
\begin{equation}
\chi=\begin{cases}
\left|\psi_x\right\rangle  & \mbox{with probability 1/2}\\
\left|\psi_{-x}\right\rangle  & \mbox{with probability 1/2}.
\end{cases}
\end{equation}
On the other hand, if Alice measured her spins along $\left\{ \left|\rightarrow\right\rangle ,\left|\leftarrow\right\rangle \right\} $ basis, then Bob will be left with the mixture
\begin{equation}
\kappa=\begin{cases}
\left|+\right\rangle  & \mbox{with probability 1/2}\\
\left|-\right\rangle  & \mbox{with probability 1/2}.
\end{cases}
\end{equation}
In standard quantum mechanics, both mixtures would be described with the density matrix 
\begin{eqnarray}
\rho & = & \frac{1}{2}\left|\psi_x\right\rangle \left\langle \psi_x\right|+\frac{1}{2}\left|\psi_{-x}\right\rangle \left\langle \psi_{-x}\right|\\
 & = & \frac{1}{2}\left|+\right\rangle \left\langle +\right|+\frac{1}{2}\left|-\right\rangle \left\langle -\right|.
\end{eqnarray}
However, under the Schroedinger-Newton theory, it is wrong to use $\rho$ because under time evolution both mixtures will evolve differently. Indeed, under time evolution driven by \eqref[name=Eq.~]{evolutionCMsn} (which has a nonlinearity of $\left\langle \hat{x}\right\rangle $) over an infinitesimal period $dt$, $\chi$ and $\kappa$ no longer remain equivalent because $\left\langle \pm|\hat{x}|\pm\right\rangle =0$, and so $\kappa$ is unaffected by the nonlinearity.

For this reason, we will have to fall back to providing probability distributions for the bath's quantum  state.  For a Gaussian state, there are many ways of doing so, as is for example shown in \figref[name=Fig.~]{densityFig}. Since this distribution likely has a large classical component (as we argue for  in the next section), we will approach the issue of thermal fluctuations by separating out contributions to thermal noise from classical and quantum uncertainty.

\begin{figure}
\centering\includegraphics[scale=0.8]{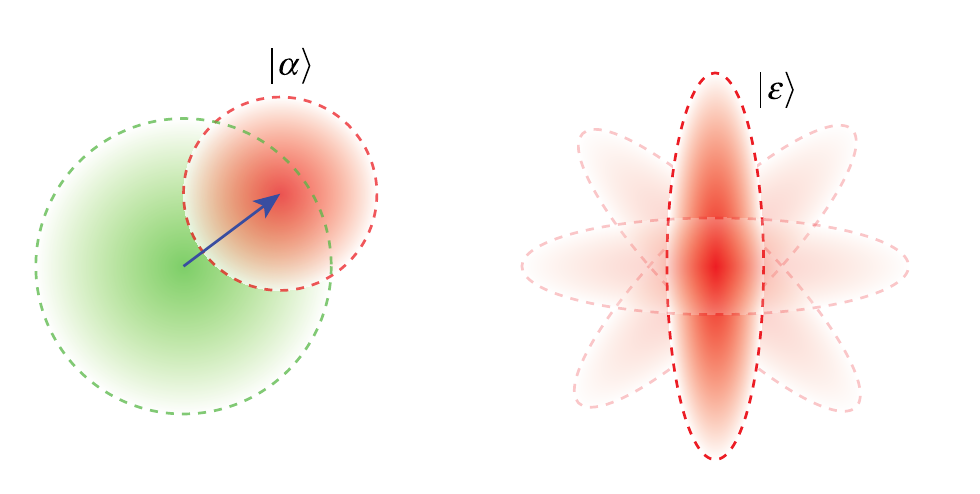}

\caption{\label{fig:densityFig} Two ways of forming the same Gaussian density matrix. In the left panel, we have an ensemble of coherent states parameterized by a complex amplitude $\alpha$, which is Gaussian distributed. The red circle depicts the noise ellipse, in phase space, of one such state.  The green ellipse depicts the total noise ellipse of the density matrix. In the right panel, we have an ensemble of squeezed states with amplitudes $\varepsilon$, which achieves the same density matrix with a fixed squeeze amplitude and a uniform distribution of squeeze angles.
}

\end{figure}

\subsection{Quantum versus classical uncertainty}

\subsubsection{Standard Quantum Statistical Mechanics}

Let us consider a damped harmonic oscillator in standard quantum mechanics, which satisfies an equation of motion of 
\begin{equation}
\label{eq:viscousdamping}
M(\ddot {\hat x}  +\gamma_m \dot {\hat x} -\ocm^2) = \Ftotth{t},
\end{equation} 
where $\gamma_m$ is the oscillator's damping rate and $\Ftotth{t}$ a fluctuating thermal force.
We have assumed {\it viscous damping}. Other forms of damping, such as structural damping, where the retarding friction force is proportional to displacement instead of velocity \cite{dampingBook},
would reduce the classical thermal noise (which will be precisely defined later in this section) at $\oq$, making the experiment easier to perform.

At a temperature $\Temp \gg \hbar\ocm/k_B$, which accurately describes our proposed setup with a test resonant frequency under a Hz, the thermal force mainly consists of classical fluctuations. We obtain $\Ftotth{t}$'s spectrum from the fluctuation-dissipation theorem,  
\begin{equation}
\label{FDT}
S_{\Ftotthi,\Ftotthi}(\Omega)= 2 \hbar
\left[
\frac{1}{\displaystyle e^{\frac{\hbar\Omega}{k_B \Temp}}-1}
+\frac{1}{2}
\right]
\frac{\mathrm{Im}[\Gc(\Omega)]}{|\Gc(\Omega)|^2} \,,
\end{equation} 
where $\Gc(\Omega)$ is the response function of $\hat x$ to the driving force $\Ftotth{t}$,
\begin{equation}
\colmagenta{G_{c}\left(\Omega\right)}=\frac{1}{M\left(\ocm^{2}-\Omega\left(\Omega+i\gamma_{m}\right)\right)},
\label{eq:defGc}
\end{equation}
and $S_{\Ftotthi,\Ftotthi}(\Omega)$ is defined by
\begin{equation}
\langle \Ftotth{\Omega} \Ftotthi^\dagger(\Omega') \rangle_{\rm sym} =S_{\Ftotthi,\Ftotthi}(\Omega)2\pi\delta(\Omega-\Omega')
\end{equation}
 with
\begin{equation}
 \langle \hat A\hat B\rangle_{\rm \colmagenta{sym}}\equiv  \frac{\langle \hat A \hat B+\hat B\hat A\rangle}{2}.
\end{equation}
Note that we have chosen a ``double-sided convention'' for calculating spectra.   

The fact that the motion of the test mass is damped due to its interaction with the heat bath also requires that the thermal force has a (usually small but nevertheless conceptually crucial) quantum component, 
\begin{equation}
\left[\hat F_{\rm th}(t),\hat F_{\rm th}(t') \right] \neq 0\,.
\end{equation}
which compensates for the decay of the oscillator's canonical commutation relations due to adding damping in its equations of  motion (refer to section 5.5 of \cite{barnett2002methods} for details). Note that the second term in the bracket in Eq.~(\ref{FDT}) provides the zero-point fluctuations of the oscillator as $T \rightarrow 0$.

\subsubsection{Quantum Uncertainty}

Let the bath be in some quantum state $\left | \Phi_B \right \rangle$ over which we will take {\it expectation values}.  The thermal force operator acting on the system can then be conveniently decomposed into
\begin{equation}
\Ftotth{t} = \fth{t} + \fzp{t}
\end{equation}
where we define
\begin{equation}
\fth{t} = \langle \Ftotth{t} \rangle\,,\quad \fzp{t} = \Ftotth{t} -\langle \Ftotth{t} \rangle.
\end{equation}
 We use the subscripts ``cl'' and ``zp'' because $\fth{t}$ is a complex number, while $\fzp{t}$ will be later chosen to drive the ``zero-point'' quantum fluctuation of the mass.

For any operator $\hat A$, we shall refer to $\langle \hat A\rangle$ as the quantum expectation value and 
\begin{equation}
V[\hat A] \equiv \langle \hat A^2\rangle -\langle \hat A\rangle^2
 \end{equation}
 as its {\it quantum uncertainty}. We also define the quantum covariance by
 \begin{equation}
 \mathrm{Cov}[\hat A,\hat B] = \langle \hat A\hat B\rangle_{\rm sym} -\langle  \hat A\rangle\langle \hat B\rangle
 \end{equation}

 Suppose $|\Phi_B\rangle$ is a Gaussian quantum  state, an assumption satisfied by  harmonic heat-baths under general conditions \cite{ShapiroDecoherence}, then $|\Phi_B\rangle$ is completely quantified by the following moments: the means
 \begin{equation}
 \langle \fth{t}\rangle = \fth{t}\,,\quad  \langle  \fzp{t} \rangle =0,
 \end{equation}
 the covariances that include $\fth{t}$
\begin{equation}
\mathrm{Cov}\left[\fth{t},\fth{t'} \right]=
\mathrm{Cov}\left[\fth{t},\fzp{t'}\right]
=0,
\end{equation}
and those that don't
\begin{eqnarray*}
\mathrm{Cov}\left[\Ftotth{t},\Ftotth{t'}\right] & = &
\mathrm{Cov}\left[\fzp{t},\fzp{t'}\right]\nonumber \\
&=&\langle \fzp{t}\fzp{t'}\rangle_{\rm sym} \neq 0.
\end{eqnarray*}

\subsubsection{Classical Uncertainty}

The state $|\Phi_B\rangle$ is drawn from an ensemble with a probability distribution $p(|\Phi_B\rangle)$.  For each member of the ensemble, we will have a different quantum expectation $\fth{t}$, and a different two-time quantum covariance for $\fzp{t}$. We shall call the variations in these  quantities {\it classical fluctuations}, because they are due to our lack of knowledge about a system's wavefunction. 

The {\it total covariance} of the thermal force, using our terminology, is given by:
\begin{align}
&\overline{\left\langle \frac{\hat F_{\rm th}(t)\hat F_{\rm th}(t')+
\hat F_{\rm th}(t')\hat F_{\rm th}(t)}{2}\right\rangle }\nonumber\\
=&
\overline{\langle\hat f_{\rm zp}(t)\hat f_{\rm zp}(t')\rangle_{\rm sym} }+ \overline{f_{\rm cl}(t)f_{\rm cl}(t')}\,,
\label{eq:totalThNoise}
\end{align}
where $\overline{\left\langle \quad\right\rangle }$ denotes taking an ensemble average over different realizations of the thermal bath.
\eqref[name=Eq.~]{totalThNoise} is the total thermal noise we obtain, and in standard quantum mechanics there is no way to separately measure quantum and classical uncertainties. 

\subsubsection{Proposed model}

We shall assume that $\hat f_{\rm zp}$'s two-time quantum covariance, $\langle\hat f_{\rm zp}(t)\hat f_{\rm zp}(t')\rangle_{\rm sym}$, provides the zero-point fluctuations in the position of the test mass, and that its ensemble average is zero (i.e. the uncertainty in $\fzp{t}$ comes solely from quantum mechanics). This results in $\fzp{t}$ having a total spectrum of:
\begin{equation}
\SthQ{\omega} = \hbar\frac{\mathrm{Im}[\Gc(\Omega)]}{|\Gc(\Omega)|^2} =\hbar \omega M \gamma_m  \,.
\label{eq:fzpSpectrum}
\end{equation}

Moreover, we shall assume that $f_{\rm cl}$'s two-time ensemble covariance,  $\overline{f_{\rm cl}(t)f_{\rm cl}(t')}$,  provides the fluctuations predicted by classical statistical mechanics.  This results in $f_{\rm cl}$ having a total spectrum of
\begin{equation}
S_{f_{\rm cl}}(\Omega) = \frac{2\hbar}{\displaystyle e^{\frac{\hbar\Omega}{k_B \Temp}}-1}\frac{\mathrm{Im}[\Gc(\Omega)]}{|\Gc(\Omega)|^2} \approx 2 k_B T M \gamma_m \,.
\label{eq:fclSpectrum}
\end{equation}

\subsection{Validity of the quadratic SN equation \label{subsec:validity}}

In general, the center of mass wavefunction $|\psi \rangle$ follows the SN equation
\begin{equation}
i\hbar\frac{d|\psi\rangle}{dt} =  \left[\hat H_{\rm NG} + \hat V\right]|\psi\rangle, 
\label{eq:SNHamiltNothNoise}
\end{equation}
where the gravitational potential $\hat{V}$ can be approximately calculated by taking an expectation value of Eq. (8) in \cite{HuanPaper} with respect to the internal degrees of freedom's wavefunction:
\begin{equation}
\hat V= \int  \mathcal{E}(\hat x-z) \left| \langle  \psi |z\rangle \right|^2 \,dz
\label{eq:Vdef}
\end{equation} 
with $\mathcal{E}$ the ``self energy'' between a shifted version of the object and itself at the original position. We calculate $ \mathcal{E}$ to be 
\begin{eqnarray*}
\mathcal{E}(x) & = & GMm\left(\frac{1}{\dxzp}-\frac{1}{x}\mbox{erf}\left(\frac{x}{2\dxzp}\right)\right)\\
 & = & \frac{GMm}{\sqrt{\pi}\dxzp}\left(\sqrt{\pi}-1+\frac{x^{2}}{12\dxzp^{2}}-\frac{x^{4}}{160\dxzp^{4}}+...\right)
\end{eqnarray*}
As a result, $\hat V$ is in general difficult to evaluate because it depends on an infinite number of expectation values. When the center of mass spread
\begin{equation}
\dxcm \equiv\sqrt{\left\langle \left(\hat{x}\left(t\right)-\left\langle \hat{x}\left(t\right)\right\rangle \right)^{2}\right\rangle }
\label{eq:dxcmDef}
\end{equation}
is much less than $\dxzp$, $\mathcal{E}$ can be approximated to quadratic order in $x$, leading to the simple quadratic Hamiltonian presented in \eqref[name=Eq.~]{evolutionCMsnintro} \cite{HuanPaper}. 
In this section, we show that classical thermal noise does not affect the condition $\dxcm \ll \dxzp$.

We include classical thermal noise in our analysis through the following interaction term:
\begin{equation}
\hat{V}_{{\rm cl}}\left(t\right)\equiv -\fth t\hat{x}.
\end{equation}
We will show that $\dxcm$ does not depend on $\fth t$, even when we use the full expression for $\hat{V}$. 

We first momentarily ignore $\hat{V}$, and show that under the non-gravitational Hamiltonian, $\hat{H}_{\rm NG}$, $\dxcm$ is unaffected by $\fth t$. Since $\hat{H}_{\rm NG}$ is quadratic, then the time-evolved position operator under $\hat{H}_{\rm NG}$, $\hat{x}^{(0)}$, is of linear form
\begin{eqnarray}
\hat{x}^{(0)}\left(t\right) & = & \sum_{i}\left(c_{i}\left(t\right)\hat{q}_{i}+d_{i}\left(t\right)\hat{k}_{i}\right) + \int\Gc\left(t-z\right)\fth{z}dz \nonumber \\ 
& & + \int r\left(t,z\right)\hat{a}_{1}\left(z\right)dz +\int s\left(t,z\right)\hat{a}_{2}\left(z\right)dz,
 \label{eq:x0Exp}
\end{eqnarray}
where $\hat{q}_{i}$ and $\hat{k}_{i}$ are canonically conjugate operators of discrete degrees of freedom such as the center of mass mode of the test mass, $\Gc\left(t\right)$ is the inverse Fourier transform of the response function defined by \eqref[name=Eq.~]{defGc}, and $r(t)$ and $s(t)$ are $c$-number functions. As a result, the variance of $\hat{x}^{(0)}$ is unaffected by $\fth t$.

The full time-evolved position operator (in the state-dependent Heisenberg picture introduced in section II.B), can be expressed in terms of $\hat{x}^{(0)}$  in the following way:
\begin{equation}
\hat{x}_H\left(t\right)=\hat{U}_{I}^{\dagger}\left(t\right)\hat{x}^{\left(0\right)}\left(t\right)\hat{U}_{I}\left(t\right),
\label{eq:xInteraction}
\end{equation}
where $\hat{U}_{I}$ is the (state-dependent) interaction picture time-evolution operator associated with
\begin{equation}
\hat{V}_{I}(t)=\hat{U}_{{\rm NG}}^{\dagger}(t)\hat{V}(t)\hat{U}_{{\rm NG}}(t). 
\end{equation}
Specifically, $\hat{U}_I$ is defined by 
\begin{equation}
\hat{U} = \hat{U}_{\rm NG} \hat{U}_I,
\end{equation}
where $\hat{U}_{\rm NG}$ is the time-evolution operator associated with $\hat{H}_{\rm NG}$.
We will show that $\hat{V}_{I}$ and $\hat{U}_{I}$ are independent of $\fth t$. 

We begin the proof, of $\hat{V}_{I}$ independent of $\fth t$, by conveniently rewriting $\left|\left\langle \psi|z\right\rangle \right|^{2}$ in \eqref[name=Eq.~]{Vdef} as the expectation value of an operator. We do so by expressing the projection $\left|z\right\rangle \left\langle z\right|$ as a delta function:
\begin{equation}
\hat{V}=\int\mathcal{E}\left(\hat{x}-z\right)\left\langle \delta\left(\hat{x}-z\right)\right\rangle dz.
\end{equation}
We then express $\mathcal{E}$ and $\delta\left(\hat{x}-z\right)$ in the Fourier domain:
\begin{equation}
\hat{V}\propto\int\mathcal{F}\left(l\right)e^{-il\left(\hat{x}-z\right)}\left\langle e^{-ik\left(\hat{x}-z\right)}\right\rangle dk \, dl \, dz,
\end{equation}
where $\mathcal{F}$ is the Fourier transform of $\mathcal{E}$.
Finally, we perform the integral over $z$, obtaining
\begin{equation}
\hat{V}\propto\int\mathcal{F}\left(k\right)e^{-ik\hat{x}}\left\langle e^{ik\hat{x}}\right\rangle dk.
\end{equation}

In the interaction picture, 
\begin{eqnarray}
\hat{V}_{I}\left(t\right) & \propto & \int\mathcal{F}\left(k\right)e^{-ik\hat{x}^{\left(0\right)}\left(t\right)}\left\langle \Psi_{0}|e^{ik\hat{x}_{H}\left(t\right)}|\Psi_{0}\right\rangle dk\nonumber \\
 & \propto & \int\mathcal{F}\left(k\right)e^{-ik\hat{x}^{\left(0\right)}\left(t\right)}\times\nonumber \\
 &  & \braOket{\Psi_{0}}{\hat{U}_{I}^{\dagger}(t)e^{ik\hat{x}^{\left(0\right)}\left(t\right)}\hat{U}_{I}(t)}{\Psi_0} dk,
 \label{eq:VIFourierExpr}
\end{eqnarray}
where $\left|\Psi_{0}\right\rangle $ is the initial wavefunction of the entire system. Notice that the linear dependence of $\hat{x}^{(0)}$ on $\fth t$ cancels out in \eqref[name=Eq.~]{VIFourierExpr}. However, $\hat{V}_{I}$ could still depend on $\fth t$ through 
$\hat{U}_{I}$. We will show that this is not the case. 

The operator 
\begin{equation}
\hat{V}_{I}\left(0\right)=\hat{V}
\label{eq:VI0eq}
\end{equation}
and the ket 
\begin{equation}
\hat{U}_{I}\left(0\right)\left|\Psi_{0}\right\rangle =\left|\Psi_{0}\right\rangle
\label{eq:UI0eq}
\end{equation}
do not depend on $\fth t$ at the initial time $t=0$. At later times,
$\fth t$ can only appear through the increments $d\hat{V}_{I}/dt$ or
$d\hat{U}_{I}\left|\Psi_{0}\right\rangle /dt$. The latter is given
by 
\begin{equation}
i\hbar\frac{d}{dt}\hat{U}_{I}\left|\Psi_{0}\right\rangle =\hat{V}_{I}\hat{U}_{I}\left|\Psi_{0}\right\rangle ,\label{eq:UIDeq}
\end{equation}
while
\begin{eqnarray}
i\hbar\frac{d\hat{V}_{I}\left(t\right)}{dt} & = & \int dk\mathcal{F}\left(k\right)\Bigg(\left[e^{-ik\hat{x}^{\left(0\right)}},\hat{H}_{{\rm NG}}\right]\times\nonumber \\
 &  & \hat{U}_{I}^{\dagger}\left\langle e^{ik\hat{x}^{\left(0\right)}}\right\rangle _{0}\hat{U}_{I}+e^{-ik\hat{x}^{\left(0\right)}}\times\nonumber \\
 &  & \left\langle \hat{U}_{I}^{\dagger}\left[e^{ik\hat{x}^{\left(0\right)}},\hat{H}_{{\rm NG}}+\hat{V}_{I}\right]\hat{U}_{I}\right\rangle _{0}\Bigg),\label{eq:VIDeq}
\end{eqnarray}
where the expectation values $\left\langle \quad\right\rangle _{0}$
are taken over $\left|\Psi_{0}\right\rangle $. In both terms in the sum, the
dependence of $\hat{x}^{(0)}$ on $\fth t$ cancels out, and so $\fth t$ does
not explicitly appear in the system of differential equations (\ref{eq:UIDeq})
and (\ref{eq:VIDeq}). $\fth t$ does not also appear in the initial conditions (\ref{eq:UI0eq}) and (\ref{eq:VIDeq}). Consequently, both
$\hat{V}_{I}$ and $\hat{U}_{I}$ are independent of $\fth t$. 

We then use \eqref[name=Eq.~]{xInteraction} to establish that the center of mass position operator is independent of $\fth t$. As a result, the exact expression for $\dxcm$ is also independent of $\fth t$. If $\dxcm \ll \dxzp$ holds in the absence of classical thermal noise, it also holds in the presence of it. We will have to check this assumption  in order for the linear Heisenberg equation to hold.  Otherwise, if $\dxcm$ becomes larger than $\dxzp$, the effect of $\hat{V}$ becomes weaker, because $\hat{V}$ becomes shallower than the quadratic potential 
$$\frac{1}{2} M \osn^2 \left( \hat x- \langle \hat{x} \rangle \right)^2. $$

\subsection{Heisenberg equations of motion with thermal noise included}

The dynamics of our proposed model for an open optomechanical system are summarized by the following state-dependent Heisenberg equations:
\begin{align}
\label{heis1}
\frac{d\hat x}{dt}  =&\frac{\hat{p}}{M}\\
\frac{d\hat p }{dt}  =&-M\ocm^{2}\hat{x}- \gamma_m \hat p - M\omega_{\rm SN}^2(\hat x -  \langle \hat x \rangle) \nonumber\\ +&\alpha\hat{a}_{1} +f_{\rm cl} + \hat f_{\rm zp}\\
\hat{b}_{1}  =&\hat{a}_{1} \\
\hat{b}_{2}=&\hat{a}_{2}+\frac{\alpha}{\hbar}\hat{x},
\label{heis4}
\end{align}
where the spectra of  $\fzp{\omega}$ and $\fth{\omega}$ are given by Eqs.  (\ref{eq:fzpSpectrum}) and (\ref{eq:fclSpectrum}), respectively. 

We solve Eqs. (\ref{heis1})--(\ref{heis4}) by working in the frequency domain, and obtain at each frequency $\omega$,
\begin{align}
\hat{b}_{2}\left(\omega\right) 
=\hat{A}\left(\omega\right)+\frac{\alpha \Gc\left(\omega\right)}{\hbar}f_{\rm cl}\left(\omega\right)+\left\langle \hat{B}\left(\omega\right)\right\rangle \label{eq:b2Dynamics}.
\end{align}
We separately discuss the three terms. The  operator $\hat{A}\left(\omega\right)$ is the linear quantum contribution to $\hat{b}_{2}$:
\begin{align}
\colmagenta{\hat{A}\left(\omega\right)}  \equiv\hat{a}_{2}\left(\omega\right)+\frac{\alpha G_{q}\left(\omega\right)}{\hbar}\left[\alpha\hat{a}_{1}+\hat{f}_{\rm zp}\left(\omega\right)\right]\,,
\label{eq:Adef}
\end{align}
where  $\hat{a}_{2}\left(\omega\right)$ represents shot noise, 
\begin{equation}
\colmagenta{G_{q}\left(\omega\right)}\equiv\frac{1}{M\left(\oq^{2}-\omega^{2}+i\omega\gamma_{m}\right)}
\end{equation}
is the quantum response function of the damped torsional pendulum's center of mass position, $\hat{x}\left(\omega\right)$, to the thermal force, and $\alpha \hat a_1$ and $\hat f_{\rm zp}$ are the quantum radiation-pressure force and the quantum piece of the thermal force acting on the test mass, respectively. 

The second term in Eq.~(\ref{eq:b2Dynamics}) represents classical thermal noise, with $\Gc\left(\omega\right)$ defined in \eqref[name=Eq.~]{defGc}. Note that the classical and quantum resonant frequencies in $\Gc({\omega})$ and $G_q({\omega})$, respectively, differ from each other.

The third term in Eq.~(\ref{eq:b2Dynamics}), $\langle \hat{B}(\omega) \rangle$, 
represents the non-linear contribution to $\hat{b}_{2}\left(\omega\right)$
\begin{equation}
\colmagenta{\hat{B}\left(\omega\right)} \equiv\frac{\alpha \Delta G\left(\omega\right)}{\hbar}\left[\alpha\hat{a}_{1}\left(\omega\right)+\hat{f}_{\rm zp}\left(\omega\right)\right],
\label{eq:Bdef}
\end{equation}
where we defined 
\begin{equation}
\colmagenta{\Delta G\left(\omega\right)}\equiv G_{c}\left(\omega\right)-G_{q}\left(\omega\right).
\end{equation}
In the next section, we discuss the subtle issue of how to convert the wavefunction average $\langle...\rangle$ to the statistics of measurement outcomes.

\section{\label{sec:Measurements-in-NLQM}Measurements in nonlinear quantum optomechanics}

With the assumption of classical gravity, we will have to revisit the wavefunction collapse postulate, because a sudden projective measurement of the outgoing optical field induces a change in the quantum state of any of its entangled partners, including possibly the macroscopic pendulum's state. As a result, we might obtain an unphysical change in the Einstein tensor which violates the Bianchi identity. Moreover, since the Schroedinger-Newton equation is nonlinear, we will show that we have to address an additional conceptual challenge: there is no unique way of extending Born's rule to nonlinear quantum mechanics. 

In this section, we propose two phenomenological prescriptions, which we term pre-selection and post-selection, for determining the statistics of an experiment within the framework of classical gravity. 

\subsection{\label{sub:Revisiting-wavefunction-collapse}Revisiting Born's rule in linear quantum mechanics}

We will use the wavefunction collapse postulate as a guide. The postulate is mathematically well defined, but can be interpreted in two equivalent ways, which become inequivalent in nonlinear quantum mechanics.

The first interpretation is widely used, and describes a quantum measurement experiment in the following way: a preparation device initializes a system's quantum state to $|i\rangle$, which evolves for some period of time under a unitary operator, $\hat U$, to
\begin{equation}
|i\rangle \rightarrow \hat U |i\rangle\,.
\end{equation}
The system then interacts with a measurement device, which collapses the system's state into an eigenstate, $\left|f\right\rangle$, of the observable associated with that device. The probability of the collapse onto $\left|f\right\rangle$ is 
\begin{equation}
p_{i\rightarrow f} \equiv  |\langle f|\hat U |i\rangle|^2. \label{eq:preBorn}
\end{equation}
We will refer to this expression of Born's rule as \emph{pre-selection}. 

Second, the unitarity of quantum mechanics allows us to rewrite
\eqref[name=Eq.~]{preBorn} to 
\begin{equation}
p_{i\rightarrow f} = |\langle i|\hat U^\dagger  |f\rangle|^2 \equiv p_{i\leftarrow f}.
\label{eq:postBorn}
\end{equation}
Interpreting this expression from right to left, as we did for \eqref[name=Eq.~]{preBorn},
we can form an alternate, although unfamiliar, narrative:  $\left|f\right\rangle $ evolves backwards
in time to $\hat{U}^{\dagger}\left|f\right\rangle $, and is then projected by the preparation device
to the state $\left|i\right\rangle $, as is illustrated in \figref[name=Fig.~]{qmWFCnarratives}. We will refer to the formulation of Born's rule based on $p_{i \leftarrow f}$ as \emph{post-selection}.

\begin{figure}
\centering\includegraphics[scale=0.7]{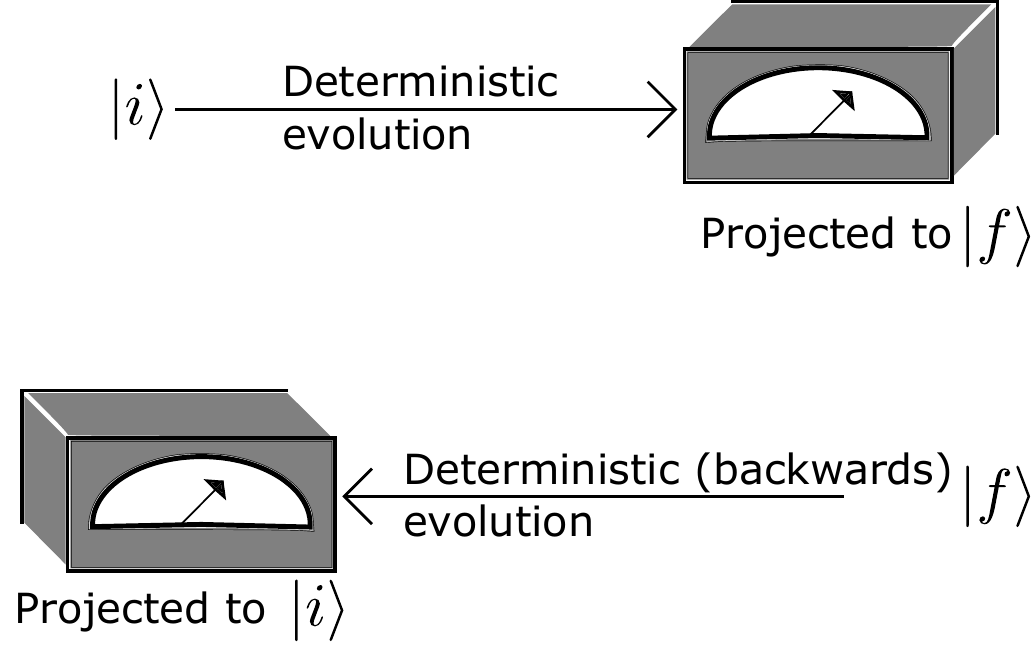}

\caption{The two prescriptions, pre-selection (top) and post-selection (bottom), that can be used to calculate measurement probabilities. Both prescriptions are equivalent in linear quantum mechanics, but become different under non-linear quantum mechanics.
\label{fig:qmWFCnarratives}}
\end{figure}

\subsection{Pre-selection and post-selection in non-linear quantum mechanics} \label{prePostSpectra}

In non-linear quantum mechanics, the Hamiltonian,
and so the time evolution operator, depends on the quantum state of the system. As a result, the pre-selection version of Born's rule, \eqref[name=Eq.~]{preBorn}, has to be revised to 
\begin{equation}
p_{i\rightarrow f} = \big|\langle f|\hat U_{|i\rangle} |i\rangle\big|^2
\end{equation}
where $\hat U_{|i\rangle}$ is the (non-linear) time evolution operator which evolves $\left| i \right\rangle$ forward in time to $\hat U_{|i\rangle} \left| i \right\rangle$.

Furthermore, the post-selection version of Born's rule, \eqref[name=Eq.~]{postBorn}, is modified to
\begin{equation}
p_{i\leftarrow f} \propto \big|\langle i |\hat U^\dagger_{|f\rangle} |f\rangle\big|^2,
\end{equation}
where $\hat U^\dagger_{|f\rangle}$
is the (non-linear) time evolution operator which evolves $\left| f \right\rangle$ backwards in time to $\hat U^\dagger_{|f\rangle} \left| f \right\rangle$. The evolution can still be interpreted as running backwards in time, because the non-linear Hamiltonians we are working with, such as in \eqref[name=Eq.~]{evolutionCMsn}, are Hermitian. Moreover, the proportionality sign follows from $$\sum_{f}\big|\langle i|\hat{U}_{|f\rangle}^{\dagger}|f\rangle\big|^{2}$$ being not, in general, normalized to unity.

Notice that $p_{i\leftarrow f}$ and $p_{f\leftarrow i}$ are in general different.  Consequently, in non-linear quantum mechanics, we can no longer equate the pre-selection and post-selection prescriptions, and we will have to consider both separately.

\subsection{Pre-selection and post-selection in non-linear quantum optomechanics \label{sub:prePostNLQM}}

In our proposed optomechanical setup, the state $|i\rangle$ is a separable state consisting of the initial state of the test object, and a coherent state of the incoming optical field, which has been displaced to vacuum, $|0\rangle_{\rm in}$ by the transformation $\hat{a}_{1,2}\rightarrow \delta\hat{a}_{1,2}+\left\langle \hat{a}_{1,2}\right\rangle$. In the pre-selection measurement prescription, as we reach steady state, the test-mass' initial state becomes irrelevant, and the system's state is fully determined by the incoming optical state.  

The set of possible states $|f\rangle$ are eigenstates of the field quadrature $b_2(t)$, which can be labeled by a time series 
\begin{equation} 
|\xi\rangle_{\rm out} \equiv |\{ \xi(t):-\infty < t < +\infty \}\rangle_{\rm out}.
\label{eq:xiOutDef}
\end{equation}
Similarly to what we discussed for pre-selection, as we reach steady state, the test-mass' initial state becomes irrelevant.  This statement can easily be demonstrated if $p_{i \leftarrow f}$ is recast in a form, {\it cf.} \eqref[name=Eq.~]{alternatePostProbForm}, where the test mass' state is forward-time evolved	and so is driven by light, and undergoes thermal dissipation.

Since $|\xi\rangle_{\rm out}$ labels a collection of Gaussian quantum states, the distribution of the measurement results $\xi(t)$ will be that of a Gaussian random process, characterized by the first and second moments. In standard quantum mechanics, they are given by the mean $\langle \hat b_2(t)\rangle$ and the correlation function 
$$\langle \hat b_2(t) \hat b_2(t')\rangle_{\rm sym} - \langle \hat b_2(t)\rangle \langle \hat b_2(t')\rangle.$$
In nonlinear quantum mechanics, the situation is subtle because $\left\langle \hat{b}_{2}\left(t\right)\right\rangle $ could depend on the measurement results  $\xi(t)$.

To determine the expression for the second moment, we will explicitly calculate $p_{i \rightarrow f}$ and $p_{i \leftarrow f}$. Since our proposed setup eventually reaches a steady state, we can simplify our analysis by working in the Fourier domain, where fluctuations at different frequencies are independent.  Note that we first ignore the classical force $\fth{t}$. We will incorporate it back into our analysis at the end of this section.

The probability of measuring $\xi$ in the pre-selection measurement prescription, 
\begin{equation}
p_{i\rightarrow f} = p_{0\rightarrow \xi} = |{}_{\rm out}\langle \xi | \hat U_{|0\rangle_{\rm in}} |0\rangle_{\rm in}|^2
\label{eq:preGeneralDef}
\end{equation}
is characterized by the spectrum of the Heisenberg Operator of $\hat b_2$ in the following way: 
\begin{equation}
p_{0\rightarrow \xi} \propto 
  \exp\left[-\frac{1}{2}\int\frac{d\Omega}{2\pi}\frac{|\xi(\Omega)-\langle \hat{b}_2(\Omega)\rangle_{0}|^2}{S_{A,A}}\right],
 \label{eq:probPre}
\end{equation}
where $\langle b_2(\Omega)\rangle_{0}$ is the quantum expectation value of the Heisenberg operator $\hat b_2(\omega)$, calculated using the state-dependent Heisenberg equations associated with an initial boundary condition of $|0\rangle_{\rm in}$, and $S_{A,A}$ is the spectral density of the linear part of $\hat{b}_2(\Omega)$, $\hat{A}$, evaluated over vacuum: 
\begin{equation*}
2\pi S_{A,A}(\omega)\delta\left(\omega-\omega^{'}\right)\equiv
\braOket{0}{\hat{A}\left(\omega\right)\hat{A}^{\dagger}\left(\omega^{'}\right)}{0}_{\rm sym}
.
\end{equation*}
Note that the derivation of \eqref[name=Eq.~]{probPre} is presented in Appendix \ref{sec:spectraDerivation}. In the same Appendix, we also show that in the limit of $\osn\rightarrow 0$, $p_{0\rightarrow\xi}$ recovers the predictions of standard quantum mechanics.

In post-section, the probability of obtaining a particular measurement record is given by
\begin{equation}
p_{i\leftarrow f} = p_{0 \leftarrow \xi}  = \left| \langle 0|\hat U^\dagger_{|\xi\rangle_{\rm out}} |\xi\rangle_{\rm out}\right|^2,
\end{equation} 
which can be written as 
\begin{equation}
p_{0 \leftarrow \xi} = 
\left| {}_{\rm out}\langle \xi|\hat U_{|\xi\rangle_{\rm out}}|0\rangle \right|^2
\label{eq:alternatePostProbForm}
\end{equation}
where 
$\hat{U}_{|\xi\rangle_{\rm out}}$
is the time-evolution operator specified by the end-state $|\xi\rangle_{\rm out}$. In Appendix \ref{sec:spectraDerivation}, we show that $p_{0\leftarrow \xi}$ is given by
\begin{equation}
\label{eq:pxipost}
p_{0 \leftarrow \xi}\propto  \exp\left[-\frac{1}{2}\int\frac{d\Omega}{2\pi}\frac{|\xi(\Omega)-\langle \hat{b}_2(\Omega)\rangle_{\xi}|^2}{S_{A,A}}\right],
\end{equation}
where $\langle \hat b_2(\Omega)\rangle_{\xi}$ is the quantum expectation value of $\hat b_2(\Omega)$'s Heisenberg operator, obtained with the state-dependent Heisenberg equations associated with the final state $|\xi\rangle$, but evaluated on the incoming vacuum state $|0\rangle$ for $\hat{a}_{1,2}$.  

Note that because $\langle b_2(\Omega)\rangle_{\xi}$ depends on $\xi$, the probability density given by Eq.~(\ref{eq:pxipost}) is modified.  We extract the inverse of the new coefficient of $|\xi^2(\Omega)|$ as the new spectrum. We will follow this procedure in \secref{signatures} C. The normalization of $p_{0 \leftarrow \xi}$ is taken care of by the Gaussian function. 

Finally, we incorporate classical noise by taking an ensemble average over different realizations of the classical thermal force, $\fthF{\omega}$. For instance, the total probability for measuring $\xi$ in pre-selection is 
\begin{equation}
\overline{p_{0\leftarrow\xi}}=\int\mathcal{D}x \; p\left(\fthF{\omega}=x\left(\omega\right)\right) \times p_{0\leftarrow\xi(x(\omega))},
\end{equation}
where $p\left(\fthF{\omega}=x\left(\omega\right)\right)$ is the probability that $f_{\rm cl}$ at frequency $\omega$ is equal to $x(\omega)$, and $\xi(x(\omega))$ is the measured eigenvalue of the observable $\hat b_2$ given that the classical thermal force is given by $x$.
The above integral can be written as a convolution and so is mathematically equivalent to the addition of Gaussian random variables. Thus, assuming independent classical and quantum uncertainties, the total noise spectrum is given by adding the thermal noise spectrum to the quantum uncertainty spectrum calculated by ignoring thermal noise.

\section{\label{sec:signatures}Signatures of classical gravity}

With a model of the bath and the pre- and post-selection prescriptions at hand, we proceed to determine how the predictions of the Schroedinger Newton theory for the spectrum of phase fluctuations of the outgoing light differ from those of standard quantum mechanics. We expect the signatures to be around $\oq$, the frequency where the Schroedinger Newton dynamics appear at, as was discussed in \secref{SN} and in \cite{HuanPaper}.

\subsection{Baseline: standard quantum mechanics}

We calculate the spectrum of phase fluctuations predicted by standard quantum mechanics, $\colmagenta{\SbQ{\omega}}$, by setting $\osn$ to 0 in \eqref[name=Eq.~]{b2Dynamics}. Making use of
\begin{equation}
S_{a_{1},a_{1}}=S_{a_{2},a_{2}}=1/2\qquad S_{a_{1},a_{2}}=0
\label{eq:vacSpectra}
\end{equation}
for vacuum fluctuations of $\hat{a}_1$ and $\hat{a}_2$, we obtain
\begin{equation}
 \SbQ{\omega} =\frac{1}{2}+\frac{\alpha^{4}}{2\hbar^{2}}\left|G_{c}\left(\omega\right)\right|^{2}+\frac{\alpha^{2}}{\hbar^{2}}\Sxth{\omega}\label{eq:sb2QM},
\end{equation}
where the first and second terms on the RHS represent shot noise and quantum radiation pressure noise  respectively, and 
\begin{equation}
\Sxth{\omega}=2k_{B} \Temp \frac{\mbox{Im}\left(G_{c}\left(\omega\right)\right)}{\omega},
\end{equation}
is the noise spectrum of the center of mass position, $\hat{x}(\omega)$, due to the classical thermal force, $\fthF{\omega}$.

We are interested in comparing standard quantum mechanics to the SN theory, which has signatures around $\oq$. Therefore, we would need to evaluate $\SbQ{\omega}$ around $\oq$. The first two terms in \eqref[name=Eq.~]{sb2QM} can be easily evaluated at $\omega=\oq$, and in the limit of $\ocm \ll \osn$, 
\begin{eqnarray}
\frac{\alpha^{2}}{\hbar^{2}}\Sxth{\omega \approx \oq} & = & \beta\Gamma^{2}\label{eq:classicalThermalAroundwq}
\end{eqnarray}
where we have defined two dimensionless quantities,
\begin{eqnarray}
\colmagenta{\beta}\equiv\frac{\alpha^{2}}{M\hbar\gamma_{m}\oq}\,,\quad \colmagenta{\Gamma^{2}} & \equiv & 2\frac{k_{B}T_{0}}{\hbar\oq}\frac{\gamma_m^{2}\oq^{2}}{\gamma_m^{2}\oq^{2}+\osn^{4}}\label{eq:Gamma2Def}.
\end{eqnarray}
$\beta$ characterizes the measurement strength (as $\alpha^2$ is proportional to the input power), and $\Gamma $ characterizes the strength of thermal fluctuations. If $Q\gg 1$, we can simplify $\Gth^2$ to
\begin{equation}
\Gth^2 \approx \frac{2k_BT_0}{\hbar\omega_{\rm SN}^3}\gamma_m^2
\label{eq:G2smallQ}.
\end{equation} 

\subsection{Signature of preselection \label{sub:sigPre}} 
In pre-selection, we evaluate the nonlinearity in \eqref[name=Eq.~]{b2Dynamics}, $\left\langle \hat{B}\left(\omega\right)\right\rangle$, over the incoming field's vacuum state, $|0\rangle_{\rm in}$:
$$\tensor[_{\rm in}]{\braOket{0}{\hat{B}(\omega)}{0}}{_{\rm in}}=0.$$
Consequently, we can directly use \eqref[name=Eq.~]{probPre} to establish that under the pre-selection measurement prescription, the noise spectrum of $\hat b_2$ is $S_{A,A}$. Taking an ensemble average over the classical force $f_{\rm cl}$ adds classical noise to the total spectrum:
\begin{eqnarray}
\SpreTot{\omega} & = & S_{A,A}\left(\omega\right)+\frac{\aopt^{2}}{\hbar^{2}}\Sxth{\omega}\label{eq:preSpectFull}.
\end{eqnarray}
Making use of \eqref[name=Eq.~]{vacSpectra}, we obtain
\begin{eqnarray}
S_{A,A}\left(\omega\right) & = & \frac{1}{2}+\SRQ{\omega}\\
\colmagenta{\SRQ{\omega}} & \equiv & \frac{\alpha^{4}}{2\hbar^{2}}\left|G_{q}\left(\omega\right)\right|^{2} + \nonumber \\
& & \; \frac{\alpha^2 \left|G_{q}\left(\omega\right)\right|^{2}}{\hbar^{2}}\SthQ{\omega}.
\end{eqnarray}
The first term in $S_{A,A}$, 1/2, is the shot noise background level, and $\SRQ{\omega}$
is the noise from quantum radiation pressure forces and quantum thermal forces. Moreover, $\SthQ{\omega}$, given by \eqref[name=Eq.~]{fzpSpectrum}, is the noise spectrum from vacuum fluctuations of the quantum thermal force $\fzp{\omega}$. 

Around $\oq$, in the narrowband limit $\gamma_m\ll\oq$, the quantum back action noise dominates and so
\begin{eqnarray*}
\SpreTot{\omega} & \approx & \left(\frac{1}{2}+\beta\Gamma^2\right) \times \nonumber
\\
& & \left[1 +\frac{\beta(\beta+2)} {2\left(1/2+\beta\Gamma^2\right)}\frac{1}{\displaystyle 1+\frac{(\omega-\oq)^2}{4\gamma_m^2}}\right].
\end{eqnarray*}
As a result, the signature of classical gravity under the pre-selection prescription can be summarized as a Lorentzian 
\begin{equation}
S(\omega) \propto 1 + \frac{h_{\rm pre}}{\displaystyle 1+4\frac{(\omega-\omega_q)^2}{\Delta_{\rm pre}^2}}
\end{equation} with a height and a full width at half maximum (FWHM) given by
\begin{equation}
h_{\rm pre} =\frac{\beta(\beta+2)}{2\left(1/2+\beta\Gamma^2\right)}\,,\quad \Delta_{\rm pre} =\gamma_m\,,
\label{eq:prehD}
\end{equation} 
respectively.
We plot the pre-selection spectrum around $\oq$ in \figref[name=Fig.~]{A-summary-of-spectra}.

\subsubsection*{\label{sub:Limits-on-pre-meas-strength}Limits on the measurement
strength}

Our results are valid only if the Schroedinger Newton potential can be approximated as a quadratic potential, which is necessary for linearizing the state-dependent Heisenberg equations, as we described in Sec.~\ref{subsec:validity}.

Specifically, we must ensure that the spread of the center of mass wavefunction excluding contributions from classical noise is significantly less than $\dxzp$, which is on the order of $10^{-11}-10^{-12}$ m for most materials (as can be determined from the discussion in section \ref{comSNEq} and Ref. \cite{Peng:zh0008}). We calculate $\dxcm$ at steady state to be
\begin{align}
\langle\hat x^2\rangle -\langle \hat x\rangle^2 & = \alpha^2  \int_{-\infty}^{+\infty} \left|G_q^2(\omega)\right|\left[\frac{1}{2}+\frac{S_{f_{\rm zp}}(\omega)}{\alpha^2}\right]\frac{d\omega}{2\pi}\nonumber\\
&  \approx  \frac{\beta+2}{2}\frac{\hbar}{2M\oq},
\label{eq:dxcmExpr}
\end{align}
where the expectation value is carried over vacuum of the input field, $|0\rangle_{\rm in}$.

\subsection{\label{sub:Post-Selection-analysis}Signature of post-selection}

In post-selection, we evaluate the nonlinearity in \eqref[name=Eq.~]{b2Dynamics}, $\left\langle \hat{B}(\omega)\right\rangle$, over the collection of eigenstates measured by the detector, $|\xi\rangle_{\rm out}$.
To determine
\begin{equation}
\colmagenta{\left\langle \hat{B}(\omega)\right\rangle _{\xi}} \equiv \tensor[_{\rm out}]{\braOket{\xi}{\hat{B}(\omega)}{\xi}}{_{\rm out}},
\end{equation}
we will make use of the fact that  $|\xi\rangle_{\rm out}$ is also an eigenstate
of $\hat{A}\left(\omega\right)$ with an eigenvalue we call
\begin{equation}
\colmagenta{\eta\left(\omega\right)} = \xi(\omega) -\left\langle \hat{B}(\omega)\right\rangle _{\xi}.
\label{eq:etaDef}
\end{equation}
The equality follows from \eqref[name=Eq.~]{b2Dynamics} with classical thermal noise ignored, which we will incorporate at the end of the calculation.
Notice that if we express $\left\langle\hat{B}(\omega)\right\rangle_\xi$ in terms of $\eta\left(\omega\right)$, we can also express it in terms of $\xi\left(\omega\right)$.

Our strategy will be to project $\hat{B}\left(t\right)$ onto the space spanned by the operators $\hat{A}\left(z\right)$ for all times $z$:
\begin{equation}
\hat{B}\left(t\right)=\int_{-\infty}^{T}\colmagenta{K\left(t-z\right)}\hat{A}\left(z\right)dz+\colmagenta{\hat{R}\left(t\right)},
\label{eq:Bproj}
\end{equation}
where $\hat{R}\left(t\right)$ is the error operator in the projection.
As a result, 
\begin{equation}
\left\langle \hat{B}\left(t\right)\right\rangle _{\xi}=\int_{-\infty}^{T}K\left(t-z\right)\eta\left(z\right)dz+\left\langle \hat{R}\left(t\right)\right\rangle _{\xi},
\end{equation}
where we made use of the definition of $\eta(t)$. In Appendix \ref{sec:Appendix-Details-of-post-calculation}, we show that if we choose $K(t)$ in such a way that $\hat{R}(t)$ and $\hat{A}(z)$ are uncorrelated for all times $t$ and $z$, 
\begin{equation}
\tensor[_{\rm in}]{\braOket{0}{\hat{R}\left(t\right)\hat{A}\left(z\right)}{0}}{_{\rm in}} + 
\tensor[_{\rm in}]{\braOket{0}{\hat{A}\left(z\right)\hat{R}\left(t\right)}{0}}{_{\rm in}} = 0
\label{eq:RAARzero}
\end{equation}
then $\left\langle \hat{R}\left(t\right)\right\rangle _{\xi}=0$. 

In the long measurement time limit, $T \gg 1$, we make use of \eqref[name=Eq.~]{Bproj} to express $\hat{R}(t)$ in terms of $\hat{B}(t)$ and $\hat{A}(z)$ and then Fourier transform \eqref[name=Eq.~]{RAARzero} to solve for $K(\omega)$. We obtain
\begin{equation}
K\left(\omega\right)=\frac{S_{B,A}\left(\omega\right)}{S_{A,A}\left(\omega\right)}
\end{equation}
Making use of \eqref[name=Eq.~]{etaDef}, we express $\left\langle \hat{B}\left(\omega\right)\right\rangle _{\xi}$ in terms of $\xi(\omega)$,
\begin{equation}
\left\langle \hat{b}_{2}\left(\omega\right)\right\rangle _{\xi}=\left\langle \hat{B}\left(\omega\right)\right\rangle _{\xi}=\frac{\xi\left(\omega\right)}{1+K\left(\omega\right)},
\end{equation}
which we then substitute into \eqref[name=Eq.~]{pxipost} to establish that post-selection's spectrum (without classical thermal noise) is given by
$$ \left|1+K\left(\omega\right)\right|^{2}S_{A,A}\left(\omega\right). $$

We finally add the contribution of classical thermal noise to $\hat b_2$'s spectrum, and obtain
\begin{equation}
\SpostTot{\omega} =\left|1+K\left(\omega\right)\right|^{2}S_{A,A}\left(\omega\right)  +\frac{\aopt^{2}}{\hbar^{2}}\Sxth{\omega}.
\end{equation}
Around $\oq$, we apply a narrowband approximation
on $\left|G_{q}\left(\omega\right)\right|^{2}$, and obtain
\begin{eqnarray}
\SpostTot{\omega \approx \oq}  \approx  \left(\frac{1}{2}+\beta\Gamma^{2}\right) \left(1+D\left(\omega\right) \right),
\label{eq:dipExpr}
\end{eqnarray}
 where 
\begin{equation*}
\colmagenta{\postSig{\omega}}  \equiv  -\frac{\beta\left(\beta+2\right)\gamma_{m}^{2}}{2 \left(1/2+\beta\Gamma^{2}\right) \left(\left(\beta+1\right)^{2}\gamma_{m}^{2}+4\left(\omega-\oq\right){}^{2}\right)}
\end{equation*}
is a Lorentzian.
By comparing $\SpostTot{\omega}$ with
$\SbQ{\omega}$, given by \eqref[name=Eq.~]{sb2QM}, we conclude that $1+\postSig{\omega}$ is the signature
of post-selection. We summarize it in the following way:
\begin{equation}
1+D(\omega) = 1 - \frac{d_{\rm post}}{\displaystyle 1+4\frac{(\omega-\oq)^2}{\Delta_{\rm post}^2}}
\end{equation}
with the depth of the dip, and its FWHM given by
\begin{equation}
d_{\rm post} =  \frac{\beta\left(\beta+2\right)}{2 \left(1/2+\beta\Gamma^{2}\right)\left(\beta+1\right)^{2}} \,,\quad \Delta_{\rm post} = \left(\beta+1\right)\gamma_m\,,
\label{eq:postParams}
\end{equation} 
respectively. A summary of the post-selection spectrum around $\oq$ is depicted in \figref[name=Fig.~]{A-summary-of-spectra}.

\begin{figure}
\centering\includegraphics[scale=0.65]{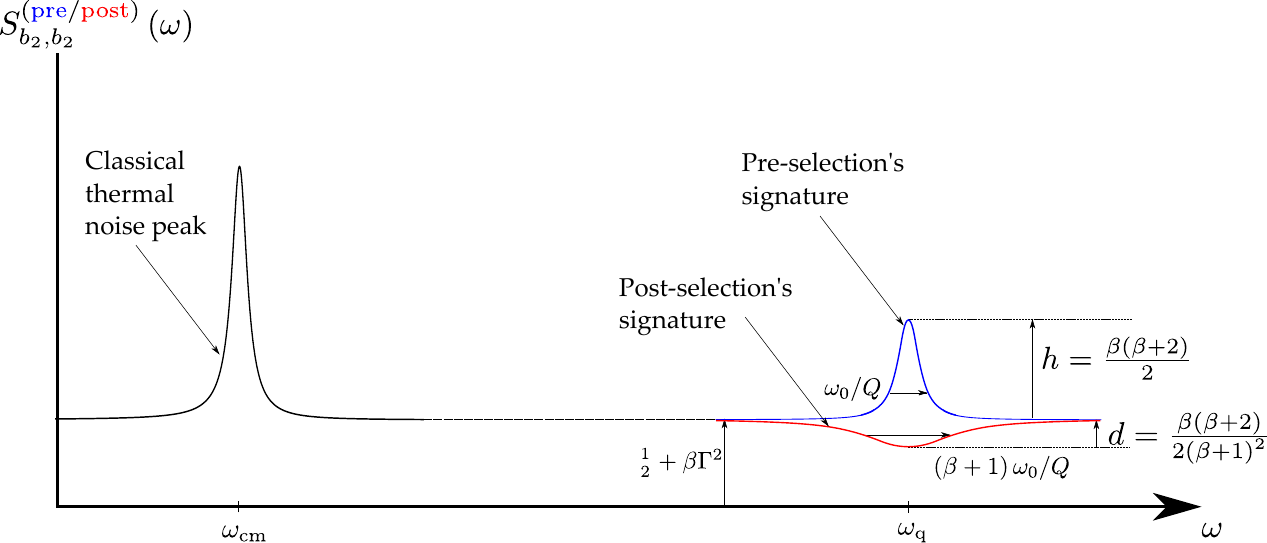}

\caption{\label{fig:A-summary-of-spectra}A depiction of the predicted signatures
of semi-classical gravity. The pre-selection measurement prescription's signature is a narrow and tall Lorentzian peak, while the post-selection measurement prescription's signature is a shallow but wide Lorentzian dip. Both prescriptions predict a Lorentzian peak of thermal noise at $\ocm$. Note that the figure is not to scale and throughout this article, we follow the convention of 2-sided spectra.}
\end{figure}

\section{\label{sec:Feasibility-analysis}Feasibility analysis}

In this section, we determine the feasibility of testing the Schroedinger-Newton theory with state of the art optomechanics setups. We will evaluate how long a particular setup would need to run for before it can differentiate between the flat noise background predicted by standard quantum mechanics around $\oq$: 
\begin{equation}
\SbQ{\omega \approx \oq}=1/2+\bMStr \Gamma^2,
\end{equation}
and the signatures of the pre- and post- measurement prescriptions, 
\begin{eqnarray*}
\SpreTot{\omega \approx \oq} & \approx &  \left(\frac{1}{2}+\beta\Gamma^{2}\right) \left( 1 + \frac{h_{\rm pre}}{\displaystyle 1+4\frac{(\omega-\omega_q)^2}{\Delta_{\rm pre}^2}} \right) \\
\SpostTot{\omega \approx \oq} & \approx &  \left(\frac{1}{2}+\beta\Gamma^{2}\right) \left(1 - \frac{d_{\rm post}}{\displaystyle 1+4\frac{(\omega-\omega_q)^2}{\Delta_{\rm post}^2}} \right),
\end{eqnarray*}
with $h_{\rm pre}$ and $\Delta_{\rm pre}$ defined by \eqref[name=Eq.~]{prehD}, and $d_{\rm post}$ and $\Delta_{\rm post}$ defined by \eqref[name=Eq.~]{postParams}.

Note that our analysis holds when the classical thermal noise peak is well resolved from the SN signatures at $\oq$. Specifically, we require that $\oq-\ocm$ be much larger than $\gamma_m$. For torsion pendulums, this is not a difficult constraint, as $\osn$ is on the order of $0.1\;\mbox{s}^{-1}$ for many materials, as is shown in Table \ref{tab:DW}.

\subsection{Likelihood ratio test}

We will perform our statistical analysis with the likelihood ratio test. Specifically, we will construct an estimator, $\estimator$, which expresses how likely the data collected during an experiment for a period $\tau$ is explained by standard quantum mechanics or the Schroedinger-Newton theory.

The estimator $\estimator$ is given by the logarithm of the ratio of the likelihood functions associated with each theory: 
\begin{equation*}
\estimator = \ln \frac{p\left(\mathcal{D}|\mbox{QM}\right)}{p\left(\mathcal{D}|\mbox{SN}\right)}
\end{equation*}
where $p\left(\mathcal{D}|\mbox{QM}\right)$ is the likelihood for measuring the data 
$$ \mathcal{D}=\left\{ \xi(t):0 <t< \tau \right\}  $$ conditioned on standard quantum mechanics being correct, and $p\left(\mathcal{D}|\mbox{SN}\right)$ is the probability of measuring the data conditioned on the Schroedinger-Newton theory, under the pre-selection or post-selection measurement prescription, being true.  Note that we will compare the predictions of standard quantum mechanics with the Schroedinger Newton theory under each prescription separately. 
All likelihood probabilities are normal distributions characterized by correlation functions which are inverse Fourier transforms of the spectra presented at the beginning of this section.

We can form a decision criterion based on $Y$. If $Y$ exceeds a given threshold, $y_{\rm th}$, we conclude that gravity is not fundamentally classical. If $Y$ is below the negative of that threshold, we conclude that the data can be explained with the Schroedinger Newton theory. Otherwise, no decision is made. 


With this strategy, we can numerically estimate how long the experiment would need to last for before a decision can be confidently made. We call this period $\tmin$ and define it to be the shortest measurement time such that there exists a threshold $y_{\rm th}$ which produces probabilities of making an incorrect decision, and of not making a decision that are both below a desired confidence level $p$.

\subsection{Numerical simulations and results}

We determined in the last section that the signatures of pre-selection and post-selection are both Lorentzians. By appropriately processing the measurement data, $\xi(t)$, the task of ruling out or validating the Schroedinger Newton theory can be reduced to determining whether fluctuations of data collected over a certain period of time is consistent with a flat or a Lorentzian spectrum centered around 0 frequency: 
\begin{equation}
S_{h (d)}\left(\omega\right)=1+\frac{h(-d)}{1+4\omega^{2}/\gamma^{2}}\quad\mbox{or}\quad S\left(\omega\right)=1
,
\end{equation}
where $\gamma$ is the full width at half maximum, $S_{h(d)}$ corresponds to a Lorentzian peak (dip) with height $h$ (depth $d$) on top of white noise.

The data can  be processed by filtering out irrelevant features except for the signatures of post- and pre-selection around $\oq$, and then shifting the spectrum: 
\begin{equation}
\tilde{\xi}(t) \equiv e^{-i \oq t}\int_{\oq-\sigma}^{\oq+\sigma}\xi\left(\Omega\right)e^{i\Omega t}d\Omega, 
\end{equation}
where $\xi(\Omega)$ is the Fourier transform of $\xi(t)$, and $\sigma$ has to be larger than the signatures' width but smaller than the separation between the classical thermal noise feature at $\ocm$ and the signatures at $\oq$.
Two independent real quadratures can then be constructed out of linear combinations of $\tilde{\xi}(t)$: 
\begin{equation}
\tilde{\xi}_c(t)  \equiv  \frac{\tilde{\xi}(t)+\tilde{\xi}^{*}(t)}{2} \,,\qquad \tilde{\xi}_s(t)  \equiv  \frac{\tilde{\xi}(t)-\tilde{\xi}^{*}(t)}{2 i}.
\end{equation}
We will carry out an analysis of the measurement time with $\tilde{\xi}_c(t)$ in mind.

We numerically generated data whose fluctuations are described by white noise, or lorentzians of different heights and depths. For example, in \figref[name=Fig.~]{histY}, we show the distribution of $Y$ for two sets of $10^5$ simulations of $\tilde{\xi}_c(t)$ over a period of  $200/\gamma$ (with $\gamma$ set to 1). In one set, $\tilde{\xi}_c(t)$ is chosen to have a spectrum of $S_d$ with $d=0.62$, and in the second set, $\tilde{\xi}_c(t)$ has a spectrum of 1. The resultant distribution for both sets is a generalized chi-squared distribution which seems approximately Gaussian. \figref[name=Fig.~]{histY} is also an example of our likelihood ratio test: if the collected measurement data's estimator satisfies $Y < - y_{\rm th}$, for $y_{\rm th}=2$, we decide that its noise power spectrum is $S_d$, if $Y > y_{\rm th}$, white noise and if $ -y_{\rm th} \leq Y \leq y_{\rm th}$, no decision is made. In \tabref{probabilities}, we show the associated probabilities of these different outcomes. Note that the choice of $y_{\rm th}$ is important, and would drastically vary the probabilities in this table.

We then determined the shortest measurement time, $\tmin$, needed to distinguish between a lotentzian spectrum and white noise, such that the probability of making a wrong decision and of not making a decision are both below a confidence level, $p$, of 10\%. Our analysis is shown in \figref[name=Fig.~]{numericalSimulations}. Since $\tilde{\xi}_c(t)$ and $\tilde{\xi}_s(t)$  are independent, we halved $\tmin$, as an identical analysis to the one performed on $\tilde{\xi}_c(t)$ can also be conducted on $\tilde{\xi}_s(t)$. 

As shown in \figref[name=Fig.~]{numericalSimulations}(a), numerical simulations of the minimum measurement time needed to decide between white noise and a spectrum of the form $S_h$, are well fitted by
\begin{equation}
\tmin(h) \approx \frac{27}{h^{0.73}} \times \frac{1}{\gamma/2},
\end{equation}
where $1/(\gamma/2)$ is the Lorentzian signature's associated coherence time. The fit breaks down for heights less than about 10. However, as we show in the next section, current experiments can easily access the regime of large peak heights.

In \figref[name=Fig.~]{numericalSimulations}(b), we show that numerical simulations of the minimum measurement time needed to decide between white noise and a spectrum of the form $S_d$, are well fitted by
\begin{equation}
\tmin(d) \approx \left( \frac{18.3}{d^{2}} - \frac{10.7}{d}  \right) \times \frac{1}{\gamma/2}.
\end{equation}
This fit is accurate, except when $d$ is close to 1. In the next section, we show that this parameter regime is of no interest to us.

Moreover, we ran simulations for higher confidence levels $p$ (in \%). We show our numerical results for pre-selection in \figref[name=Fig.~]{preConfScaling}. For $h$ between 1000 and 4000, a decrease in $p$ from 10\% to 1\% results in a 4.5-5.5 fold increase in $\tmin$. Our results for post-selection are presented in \figref[name=Fig.~]{postConfScaling}. For $d=0.62$ (which, as we show in the next section, is the normalized depth level at which most low thermal noise experiments will operate at), then $\tmin$ as a function of $p$ is well summarized by 
\begin{equation*}
\tmin(d=0.62,p) \approx \left( 2.94 - 7.38 \times \mathrm{erfc}^{-1}\left(\frac{p}{100}\right) \right)^2 \times \frac{1}{\gamma/2}.
\end{equation*}
We can also fit $\tmin(d,p)$ at other values of $d$ by a function of this form.

In the following sections, we present scaling laws for the minimum measurement time, $\tmin$, given a confidence level of 10\%, in terms of the parameters of an optomechanics experiment, and with the measurement strength $\beta$ optimized over,  for both the pre-selection and post-selection measurement prescriptions.

\begin{figure}
\centering
\includegraphics[width= 3in]{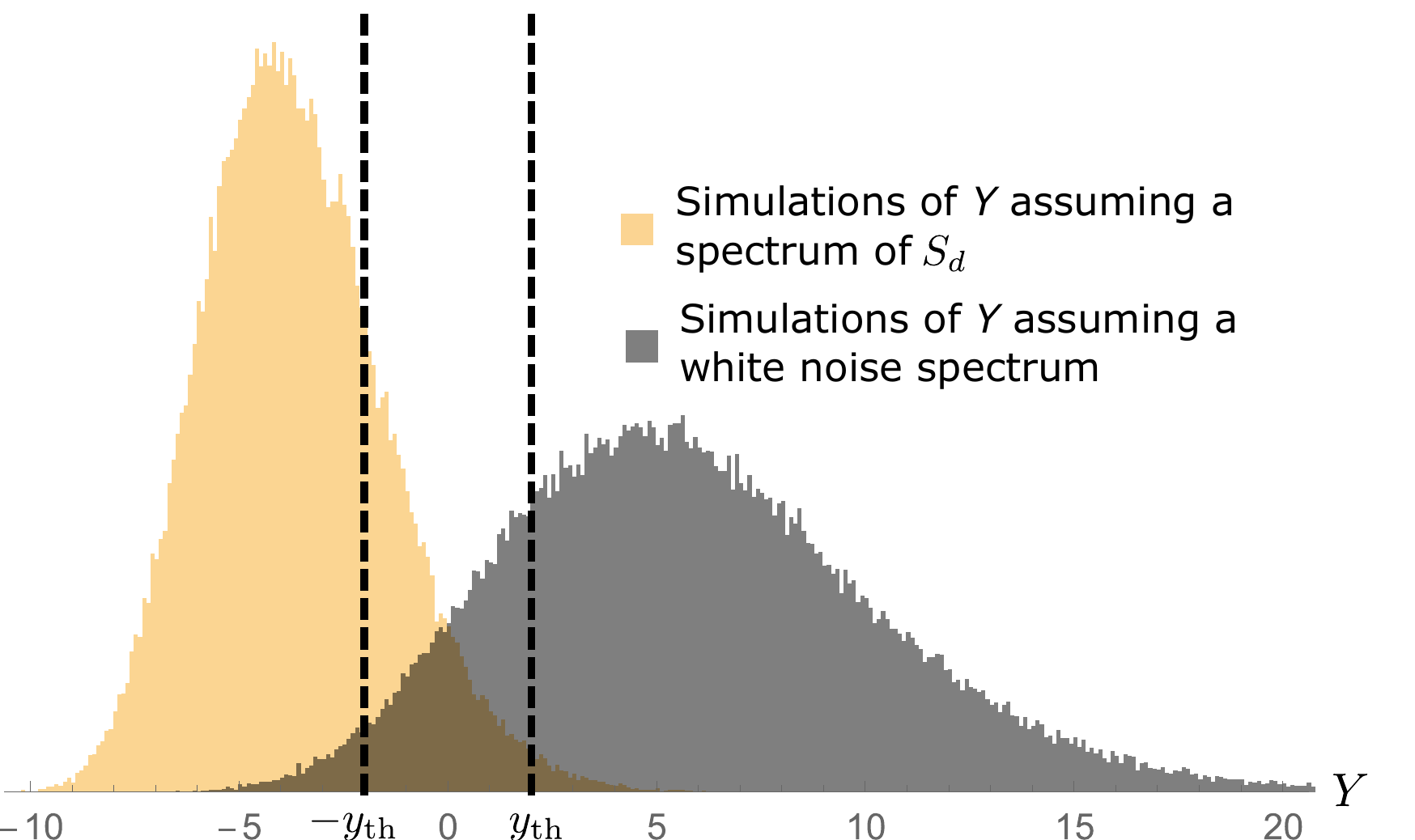}

\caption{ A histogram showing the distribution of two sets of $10^5$ realizations of $\tilde{\xi}_c(t)$ over a period of  $200/\gamma$ (with $\gamma$ set to 1), and a time discretization of $dt=0.14/\gamma$. In one set, $\tilde{\xi}_c(t)$ is chosen to have a spectrum of $S_d$ with $d=0.62$, and in the second set, $\tilde{\xi}_c(t)$ has a spectrum of 1. $y_{\rm th}$, which is chosen to be 2 in this example, allows us to construct a decision criterion: if the collected measurement data's estimator satisfies $Y < - y_{\rm th}$, we decide that its noise power spectrum is $S_d$, if $Y > y_{\rm th}$, white noise and if $ -y_{\rm th} \leq Y \leq y_{\rm th}$, no decision is made.  }
\label{fig:histY}
\end{figure}

\begin{table}

\begin{tabular}{|c|c|c|c|}
\cline{2-4} 
\multicolumn{1}{c|}{} & $\mathbb{P}\left(\mbox{correct}\right)$ & $\mathbb{P}\left(\mbox{wrong}\right)$ & $\mathbb{P}\left(\mbox{indecision}\right)$\tabularnewline
\hline 
\specialcell{Data has $S_{d}$ spectrum} & 78.7\% & 1.1\% & 20.2\%\tabularnewline
\hline 
\specialcell{Data has $S=1$ \\ spectrum} & 80.2\% & 2.1\% & 17.7\%\tabularnewline
\hline 
\end{tabular}

\caption{\label{tab:probabilities} The probabilities of the  different outcomes of the likelihood ratio test on a particular measurement data stream with an estimator following either of the two distributions shown in \figref[name=Fig.~]{histY}. The three possible  outcomes are (1) deciding that the data has a spectrum of $S_d$, (2) deciding that it has a white noise spectrum ($S=1$) or (3) making no decisions at all. $\mathbb{P}\left(\mbox{correct}\right)$ stands for the probability of deciding (1) or (2) correctly, $\mathbb{P}\left(\mbox{wrong}\right)$ is the probability of making the wrong decision on what spectrum explains the data, and $\mathbb{P}\left(\mbox{indecision}\right)$ is the probability of outcome 3. Note that a different table would have been generated if a different threshold, $y_{\rm th}$, had been chosen in \figref[name=Fig.~]{histY}.   }

\end{table}

\begin{figure}
\subfloat[Time required to distinguish a flat spectrum from a Lorentzian peak. The dashed line is a fit of $13.5/h^{0.73}$.]{\includegraphics[width= 3in]{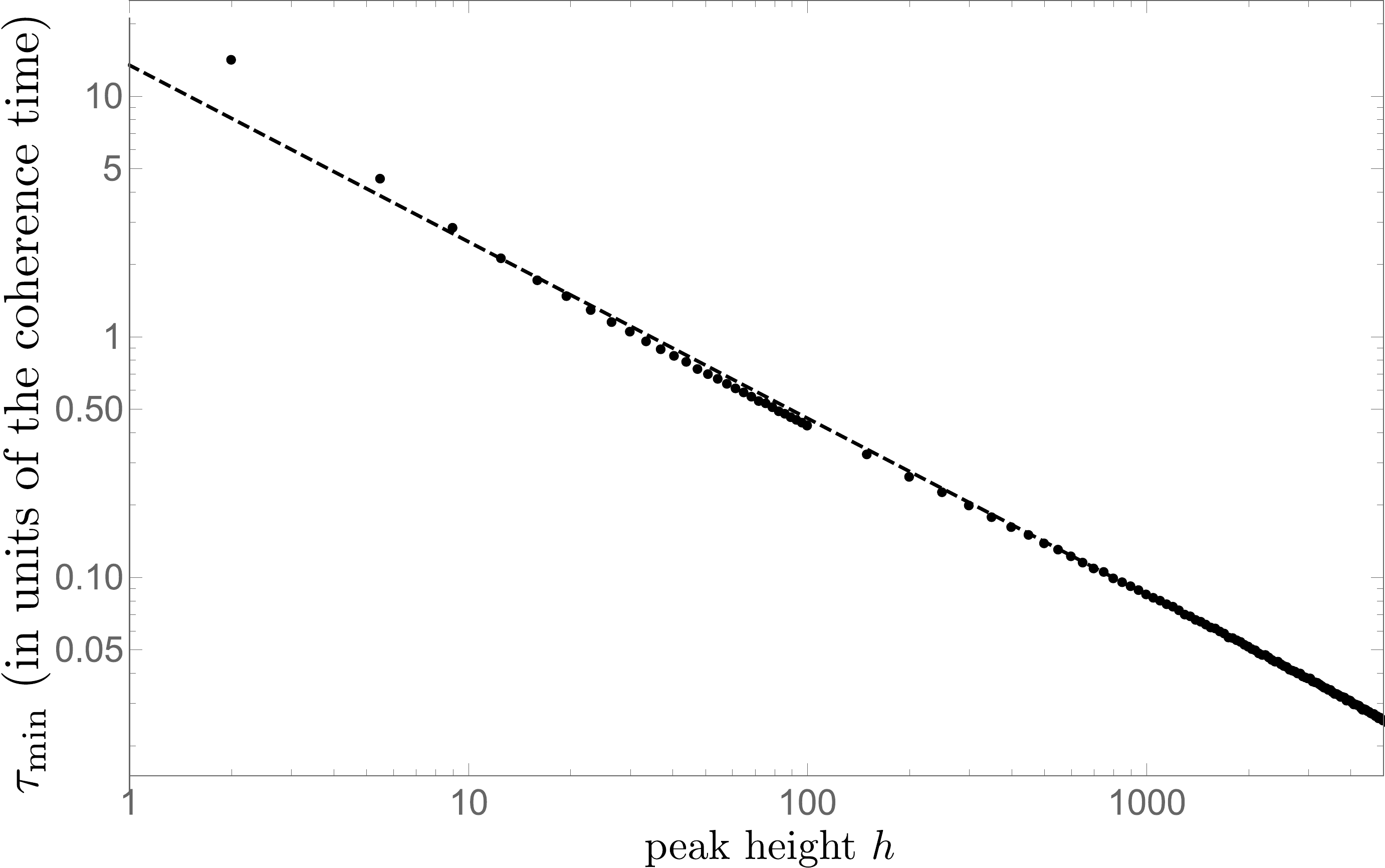}}\\
\subfloat[Time required to distinguish a flat spectrum from a Lorentzian dip. The dashed line is a fit of $18.3/d^2-10.7/d$.]{\includegraphics[width= 3in]{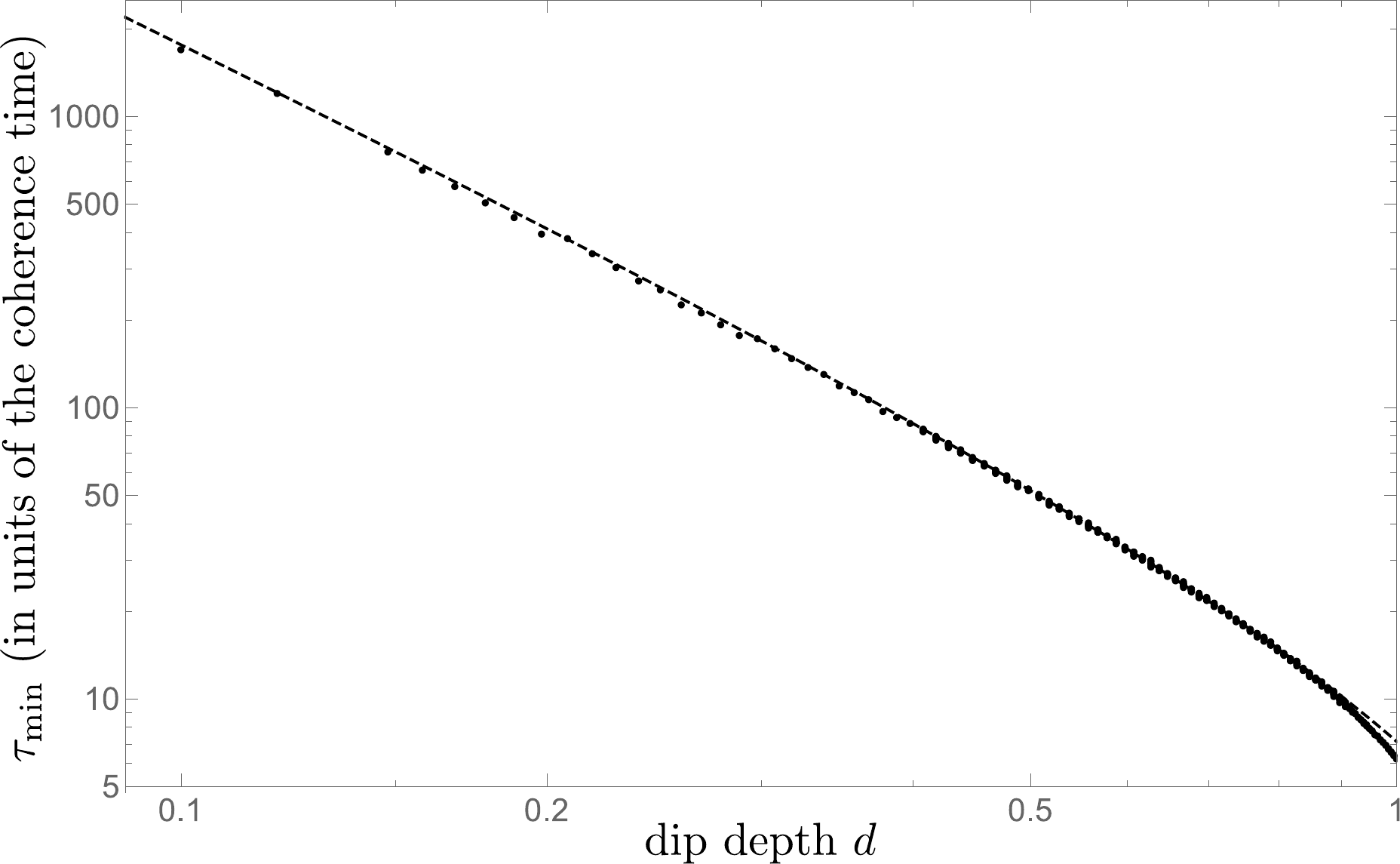}}

\caption{Simulation results showing the minimum measurement time, $\tau_{\rm min}$, required to distinguish between a Lorentzian spectrum and a flat background in such a way that the probabilities of indecision and of making an error are both below $10$\%. Plot (a) shows results for a Lorentzian peak, while plot (b) is for a Lorentzian dip. The coherence time is given by the inverse of the half width at half maximum of the Lorentzian. Note that both plots are log-log plots.} \label{fig:numericalSimulations}
\end{figure}

\begin{figure}
\centering
\includegraphics[width= 3in]{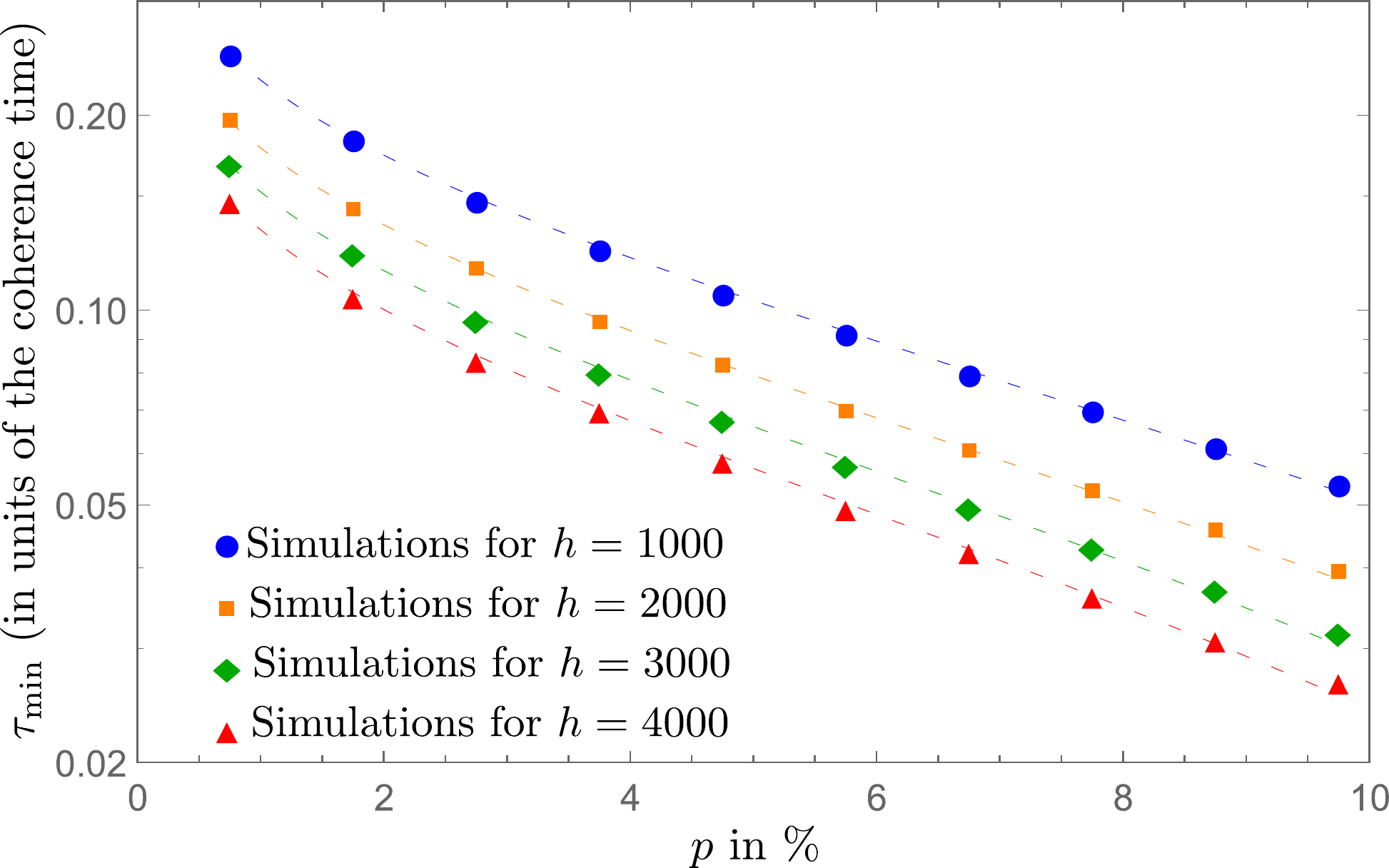}

\caption{Simulation results showing the minimum measurement time, $\tau_{\rm min}$, required to distinguish between the Schroedinger-Newton theory with the pre-selection measurement prescription (which has the signature of a Lorentzian with depth $h$) and standard quantum mechanics in such a way that the probabilities of indecision and of making an error are both below $p$\%. The coherence time is given by the inverse of the half width at half maximum of the Lorentzian. Note that the y-axis is on a log scale. Moreover, the dashed lines are \emph{only} to guide the eye (and are fits of the form $ a \; \mathrm{ln}\left( p \right)+b$). }
\label{fig:preConfScaling}
\end{figure}

\begin{figure}
\centering
\includegraphics[width= 3in]{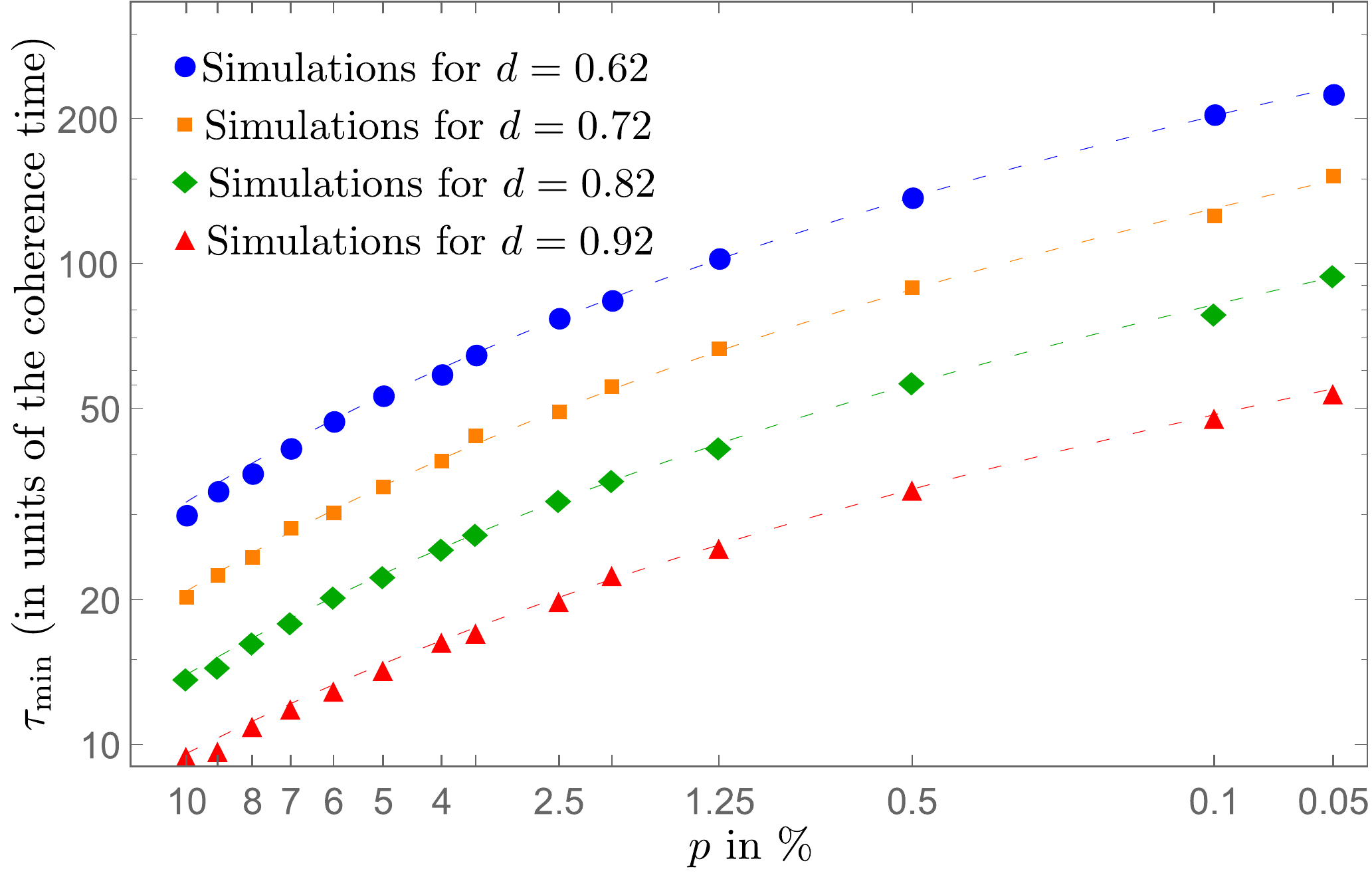}

\caption{Simulation results showing the minimum measurement time, $\tau_{\rm min}$, required to distinguish between the Schroedinger-Newton theory with the post-selection measurement prescription (which has the signature of a Lorentzian with depth $d$) and standard quantum mechanics in such a way that the probabilities of indecision and of making an error are both below $p$\%. The coherence time is given by the inverse of the half width at half maximum of the Lorentzian. Note that the x-axis is scaled by the inverse of the complimentary error function, $\rm{erfc}^{-1}$, and the y-axis is on a log scale. Moreover, the dashed lines are to guide the eye and are fits of the form $\left(a-b \times {\rm{erfc}^{-1}}\left( p/100 \right) \right)^2$. } 
\label{fig:postConfScaling}
\end{figure}

\subsection{\label{sub:Feasibility-of-pre}Time required to resolve pre-selection's signature}

The normalized pre-selection signature's height, $h_{\rm pre}$ given by \eqref[name=Eq.~]{prehD}, is a monotonically increasing function of $\bMStr$. Consequently, the larger $\beta$ is, the easier it would be to distinguish pre-selection from standard quantum mechanics. Using \eqref[name=Eq.~]{prehD} and the fit given in \figref[name=Fig.~]{numericalSimulations}a of $13.5/h^{0.73}$ (in units of the Lorentzian signature's coherence time), $\tmin$  in the limit of large $\bMStr$ scales as approximately 
\begin{equation}
\tmin \approx \frac{27}{\gamma_m} \left(\frac{2\Gth^2}{\bMStr}\right)^{0.73}.
\end{equation}

It seems that arbitrarily increasing the measurement strength would yield arbitrarily small measurement times. However, as explained in subsection \ref{sub:sigPre}, our results hold for $\dxcm \ll \dxzp$, which places a limit on $\bMStr$ of $$\bMStr\ll\frac{2\dxzp^{2}}{\hbar/\left(2M\oq\right)},$$   
where we made use of the expression for $\dxcm$ given by \eqref[name=Eq.~]{dxcmExpr}.

Placing the limit on $\bMStr$ at $1/10$ the quoted value above, for $ h \gtrsim 10 $, $\tmin$ scales with the experimental parameters in the following way: 
\begin{equation}
\begin{aligned}
\tmin  \sim 1.6 \mbox{ hours}\times\left(\frac{\Temp}{300\;{\rm K}}\right)^{0.73}\times\left(\frac{\ocm}{2 \pi \times 10 \; \rm{ mHz}}\right)^{0.47} \\
\times\left(\frac{184\;{\rm amu}}{m}\right)^{0.49}\times\left(\frac{200\;{\rm g}}{M}\right)^{0.73} \\
\times\left(\frac{10^{4}}{Q}\right)^{0.47} \times\left(\frac{0.359 \; s^{-1}}{\osn}\right)^{1.96}
\end{aligned}
\end{equation}
where $m$ is the mass of a constituent atom of the test mass, and we have assumed that the test mass is made out of Tungsten. 

Using the expressions for the measurement strength and for $\alpha^2$, given by  \eqref[name=Eq.~]{Gamma2Def} and \eqref[name=Eq.~]{alphaDef}, respectively, we determine that the input optical power needed to reach the above quoted value of $\tmin$ is 

\begin{equation}
\begin{aligned}
I_{\rm in} \approx 432 \;\mbox{mW}\times\left(\frac{10^{4}}{Q}\right)\times \left(\frac{m}{184\;{\rm amu}}\right)^{2/3}\times 
\left(\frac{M}{200\;{\rm g}}\right)^{2}\times \\ 
\left(\frac{\ocm}{2\pi\times10\;{\rm mHz}}\right) \times \left(\frac{\osn}{0.359\;s^{-1}}\right)^{2/3}\times \\ 
\left(\frac{2\pi\times 0.2\;{\rm THz}}{\omega_{c}}\right)\times\left(\frac{T}{10^{-2}}\right)^{2}.
\end{aligned}
\end{equation}


We are allowed to make use of the fit presented in \figref[name=Fig.~]{numericalSimulations}(a), of $\tmin = 27/h^{0.73}$ (in units of the coherence time), which holds only for $h \gtrsim 10$, because the pre-selection signature's normalized peak height can be easily made to satisfy this constraint. Indeed, for the parameters given above 
\begin{equation*}
\begin{aligned}
h\approx8235 \times \left(\frac{Q}{10^{4}}\right)^{2}\times\left(\frac{m}{184\;{\rm amu}}\right)^{2/3}\times\left(\frac{M}{200\;g}\right)\times \\ 
\left(\frac{2\pi\times10\;{\rm mHz}}{\ocm}\right)^{2} \times \left(\frac{\osn}{0.359\;s^{-1}}\right)^{8/3}\times\left(\frac{300\;K}{T_{0}}\right).
\end{aligned}
\end{equation*}

\subsection{\label{sub:Feasibility-of-post}Time required to resolve post-selection's signature}

As indicated by \eqref[name=Eq.~]{postParams}, the depth and width of post-selection's signature are determined by 3 parameters: $\bMStr$, $\Gth^2$ and $\gamma_m$. For a given $\Gth^2$, we can determine the optimal measurement strength $\bMStr$ that would minimize $\tmin$. We numerically carried out this analysis, and we show our results in \figref[name=Fig.~]{PostSimulationsExpt}.
For $\Gth^2$ less than about 0.1, the optimal choice of the measurement strength seems to follow a simple relationship: 
$$\bMStr_{\rm opt} \approx \frac{0.31}{\Gth^2},$$
with a corresponding measurement time, $\tmin$, of about $200\Gth^2/\gamma_m$. Note that this is a soft minimum, as large deviations from $\bMStr_{\rm opt}$ still yield near optimal values of $\tmin$. Specifically, measurement strengths roughly between $0.1/\Gth^2$ and $0.7/\Gth^2$ achieve measurement times below $225\Gth^2/\gamma_m$. 

Moreover, in the parameter regime of $\Gth^2 < 0.1$, the normalized post-selection dip depth at $\bMStr_{\rm opt}$ is 0.62, which falls well in the region where the fit presented in \figref[name=Fig.~]{numericalSimulations}(b), of $\tmin = 18.3/d^2-10.7/d$ (in units of the coherence time), is accurate.

In the limit of $\osn\gg\ocm$, the optimal measurement time scales as
\begin{equation}
\begin{aligned}
\tmin \sim \mbox{13 days}\times\left(\frac{10^{7}}{Q}\right)\times\left(\frac{T_{0}}{1\;{\rm K}}\right) \\ \times\left(\frac{0.488\;s^{-1}}{\osn}\right)^{3}\times\left(\frac{\ocm}{2 \pi \times 4 \; \rm{mhz} }\right),
\label{eq:tMinPost}
\end{aligned}
\end{equation}
where we assumed that the mechanical oscillator is made out of Osmium. Moreover, the input optical power needed to reach the above quoted value of $\tmin$ is 
\begin{equation}
\begin{aligned}
I_{\rm in}  \approx 4.8 \;\mbox{nW}\times\left(\frac{Q}{10^{7}}\right)\times\left(\frac{1\;{\rm K}}{T_{0}}\right)^{2}\times\left(\frac{M}{200\;{\rm g}}\right)^{2} \times \\
\left(\frac{2\pi\times4\;{\rm mHz}}{\ocm}\right)\times\left(\frac{\osn}{0.488\;s^{-1}}\right)^{4}\times \\ \left(\frac{2\pi\times0.2\;{\rm THz}}{\omega_{c}}\right)  
\times\left(\frac{T}{10^{-2}}\right)^{2}.
\end{aligned}
\end{equation}


Finally, we note that the experiment does not need to remain stable, or to operate, for the entire duration of $\tmin$. Since the coherence
time of the post-selection signature,
$$\frac{1}{(\beta_{\rm opt}+1)\gamma_m}, $$
is much less than $\tmin$ (in the example above, the coherence time is 5 hours), the experiment can be repeatedly run over a single coherence time. Alternatively, numerous experiments can be run in parallel.

\begin{figure}
\subfloat[$\tmin$ for different values of $\Gth^2$, $\beta$ and $\gamma_m$]{\includegraphics[width= 3in]{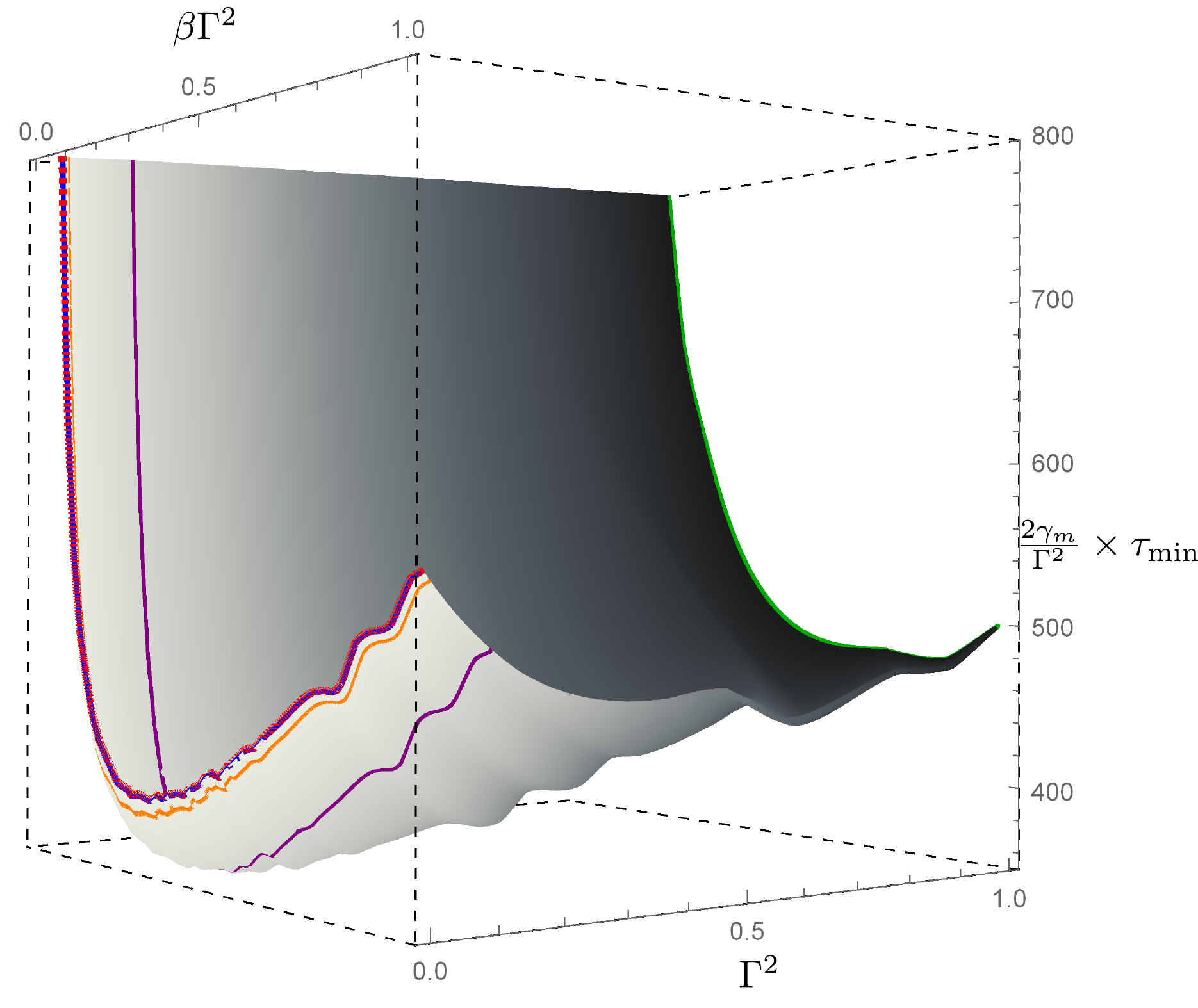}}\\
\subfloat[$\tmin$ for fixed values of $\Gth^2$]{\includegraphics[width= 3in]{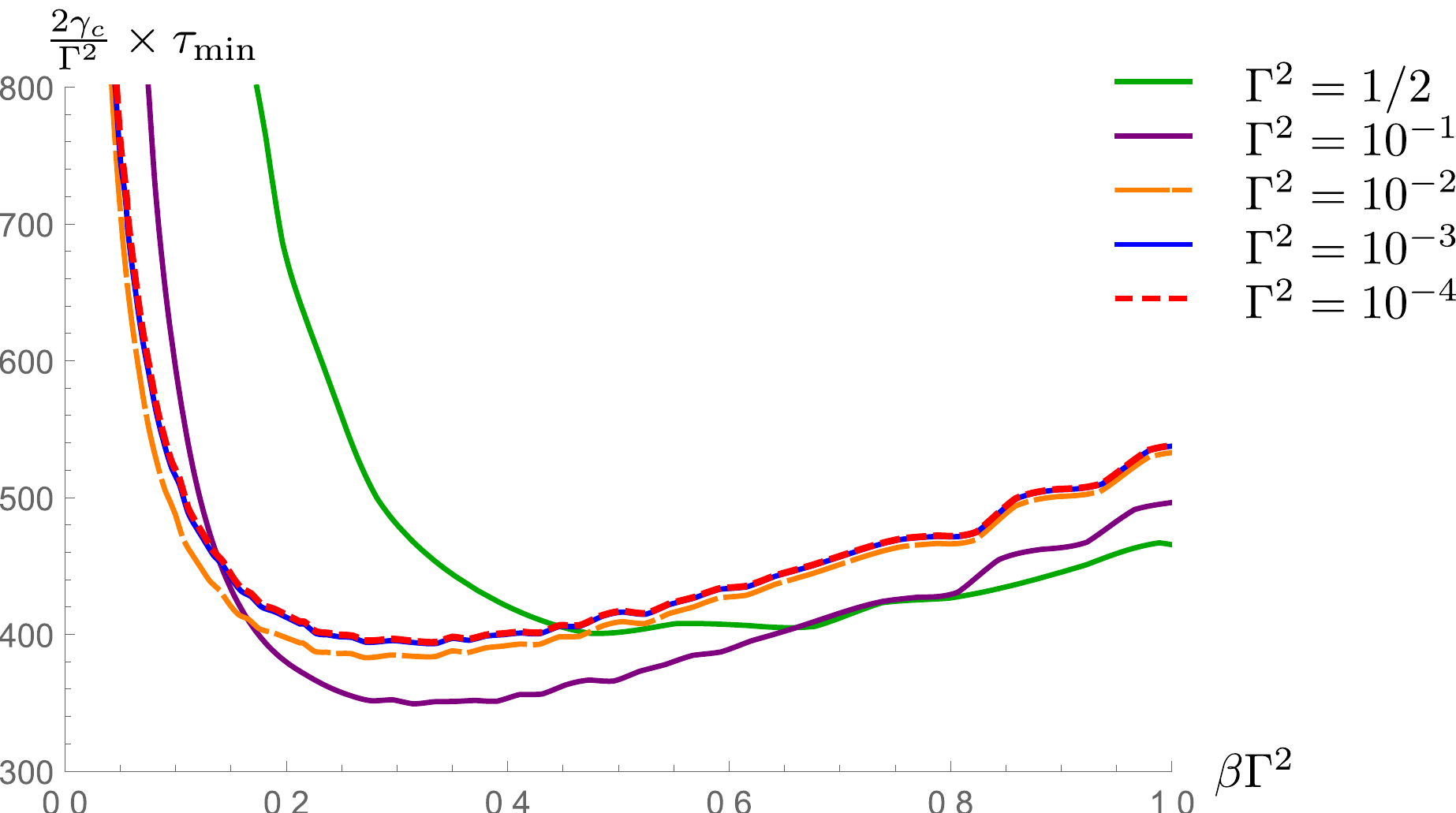}}

\caption{Minimum measurement time required to distinguish between the Schroedinger-Newton theory with the post-selection measurement prescription and standard quantum mechanics in such a way that the probabilities of indecision and of making an error are both below 10\%. Note that we interpolated the data given in \figref[name=Fig.~]{numericalSimulations} to create this figure.} \label{fig:PostSimulationsExpt}
\end{figure}

\section{\label{sec:conclusions} Conclusions }

We proposed optomechanics experiments that would look for signatures of classical gravity. This theory appreciably modifies the free unmonitored dynamics of the test mass when the following two criteria are met. First, the choice of material for the test mass is crucial. We recommend crystals with tightly bound heavy atoms around their lattice sites. Tungsten and Osmium crystals meet this criterion. Second, we recommend that the resonant frequency of the test mass be as small as possible. Torsion pendulums meet this requirement.

When adding thermal noise and measurements to our analysis, we encountered two conceptual difficulties. Both appear because the Schroedinger-Newton equation is non-linear. The first difficulty is the breakdown of the density matrix formalism. As a consequence, we had to propose a specific ensemble of pure states to describe the quantum state of the thermal bath.

The second difficulty is generalizing Born's rule to nonlinear quantum mechanics. In \secref{Measurements-in-NLQM}, we provided two prescriptions for calculating probabilities in the Schroedinger-Newton theory.  The first prescription, which we term pre-selection, takes the probability of obtaining a particular measurement result to be the modulus squared of the overlap between the forward-evolved initial state, which we choose as a boundary state for the non-linear time evolution operator, and the eigenstate corresponding to that measurement result. The second prescription, which we term post-selection, takes the probability of obtaining a particular measurement result to be the modulus squared of the overlap between the backwards-evolved measured eigenstate, which we choose as a boundary state for the non-linear evolution operator, and the initial state. Note that the predictions of both pre-selection and post-selection are consistent with that of linear quantum mechanics in the limit that the Schroedinger-Newton nonlinearity vanishes (\emph{i.e.} $\osn\rightarrow0$).

We then proceeded to obtain the signatures of classical gravity predicted by both these prescriptions in the spectrum of phase fluctuations of the outgoing light. Both signatures are Lorentzians centered around  the frequency $\oq$. The pre-selection prescription predicts a peak, while post-selection predicts a dip.
We summarize these features in \figref{A-summary-of-spectra}, which is valid when the resonant frequency of the mechanical oscillator, $\ocm$, is much smaller than $\osn$.

Finally, in the limit of the classical thermal noise peak being well separated from the SN signatures, we numerically simulated the experiment's expected measurement results and determined that pre-selection is easily testable with current optomechanics technology. However, testing post-selection will be much more challenging, although is feasible with state-of-the-art experimental parameters. In particular, we require cryogenic temperatures and a high $Q$ low frequency torsion pendulum made out of a material with a high $\osn$. \eqref[name=Eq.~]{tMinPost} contains the scaling of the minimum measurement time required to confidently test post-selection with these experimental parameters.

\begin{acknowledgments} 
We thank K.\ Thorne, J.\ Preskill, P.C.E.\ Stamp, H. Miao, Y. Ma, C. Savage, and H. Yang for discussions.  Research of Y.C.\ and H.L.\ are supported by NSF grants PHY-1404569 and PHY-1506453, as well as the Institute for Quantum Information and Matter, a Physics Frontier Center. 
\end{acknowledgments} 

\clearpage

\onecolumngrid
\appendix

\section{\label{sec:consEnergy} Conservation of energy in the SN theory}

Consider the SN equation for a collection of $N$ particles of mass
$m$:

\begin{equation}
\hat{V}_{SN}=-Gm^{2}\sum_{ij=1}^{N}\int dx_{j}\frac{p_{j}\left(x_{j}\right)}{\left|\hat{x}_{i}-x_{j}\right|}
\end{equation}
where $p_{j}\left(x_{j}\right)$ is the probability distribution for
the $j$th particle to be at location $x_{j}$:
\begin{equation}
p_{j}\left(x_{j}\right)=\int\left(\prod_{i=1}^{N}dy_{i}\right)\delta\left(y_{j}-x_{j}\right)\left|\Psi\left(y_{1},y_{2},...,y_{N}\right)\right|^{2}.
\end{equation}
$\Psi$ is the many-body wavefunction for these $N$ particles.

Let us investigate conservation of energy within the SN theory. In
standard quantum mechanics, the energy operator is given by the Hamiltonian.
Our non-linear Hamiltonian is 
\begin{equation}
\hat{H}=\sum_{i=1}^{N}\frac{\hat{P}_{i}^{2}}{2m}+\hat{V}_{\rm NG}\left(\hat{x}_{1},...,\hat{x}_{N}\right)+\hat{V}_{\rm SN},
\end{equation}
where $\hat{V}_{\rm NG}\left(\hat{x}_{1},...,\hat{x}_{N}\right)$ encodes
the non-gravitational potential energy. Under the non-linear SN theory,
$\hat{H}$ is not conserved because of $\hat{V}_{\rm SN}$'s dependence
on the wavefunction:
\begin{equation}
\frac{d\hat{H}}{dt}=\frac{\partial\hat{H}}{\partial t}=\partial_{t}\hat{V}_{\rm SN}\neq0
\end{equation}
Is there a quantity that is conserved? Consider 
\begin{equation}
\hat{E}=\sum_{i=1}^{N}\frac{\hat{P}_{i}^{2}}{2m}+\hat{V}_{\rm NG}\left(\hat{x}_{1},...,\hat{x}_{N}\right)+\beta\hat{V}_{\rm SN},
\end{equation}
where $\beta$ is to be determined such that $d\left\langle \hat{E}\right\rangle /dt=0$.
We will show that $\beta=1/2$ meets this condition. 

We begin the proof with the Heisenberg equation of motion for $\hat{E}$.
By expressing $\hat{E}$ as $\hat{H}-\left(1-\beta\right)\hat{V}_{\rm SN}$,
we obtain
\begin{eqnarray}
\frac{d\hat{E}}{dt} & = & \frac{i}{\hbar}\left[\hat{H},\hat{E}\right]+\frac{\partial\hat{E}}{\partial t}\\
 & = & \frac{i\left(1-\beta\right)Gm^{2}}{2\hbar m}\sum_{i}\sum_{jk}\int dx_{k}\left[\hat{P}_{i}^{2},\frac{p_{k}\left(x_{k}\right)}{\left|\hat{x}_{j}-x_{k}\right|}\right]-\beta Gm^{2}\sum_{ij=1}^{N}\int dx_{j}\frac{\dot{p}_{j}\left(x_{j}\right)}{\left|\hat{x}_{i}-x_{j}\right|}.\nonumber 
\end{eqnarray}
Taking the expectation value of both sides, and evaluating the commutator
in the first term, we obtain 
\begin{eqnarray*}
\frac{d\left\langle \hat{E}\right\rangle }{dt} & = & \frac{-\left(1-\beta\right)Gm}{2}\sum_{ik}\int dx_{k}\left\langle \hat{P}_{i}\frac{p_{k}\left(x_{k}\right)}{\left|\hat{x}_{i}-x_{k}\right|^{2}}+\frac{p_{k}\left(x_{k}\right)}{\left|\hat{x}_{i}-x_{k}\right|^{2}}\hat{P}_{i}\right\rangle -\beta Gm^{2}\sum_{ij=1}^{N}\int dx_{i}\int dx_{j}\frac{p_{i}\left(x_{i}\right)\dot{p}_{j}\left(x_{j}\right)}{\left|x_{i}-x_{j}\right|}.
\end{eqnarray*}
We then evaluate the expectation value in the first term. Defining the vector $\bm{x}\equiv\left(x_{1},...,x_{N}\right)$, we have
\begin{eqnarray*}
 &  & \left\langle \hat{P}_{i}\frac{p_{k}\left(y_{k}\right)}{\left|\hat{x}_{i}-y_{k}\right|^{2}}+\frac{p_{k}\left(y_{k}\right)}{\left|\hat{x}_{i}-y_{k}\right|^{2}}\hat{P}_{i}\right\rangle \\
 & = & \int d\bm{x}\Psi\left(\bm{x}\right)^{*}\left(-i\hbar\partial_{x_{i}}\frac{p_{k}\left(y_{k}\right)}{\left|x_{i}-y_{k}\right|^{2}}\Psi\left(\bm{x}\right)\right)+\int d\bm{x}\Psi\left(\bm{x}\right)^{*}\frac{p_{k}\left(y_{k}\right)}{\left|x_{i}-y_{k}\right|^{2}}\left(-i\hbar\partial_{x_{i}}\Psi\left(\bm{x}\right)\right).
\end{eqnarray*}
Next, we integrate by parts multiple times, and use that 
\begin{equation}
\frac{p_{k}\left(y_{k}\right)}{\left|x_{i}-y_{k}\right|^{2}}=-\partial_{x_{i}}\frac{p_{k}\left(y_{k}\right)}{\left|x_{i}-y_{k}\right|},
\end{equation}
 to obtain 
\begin{eqnarray*}
 &  & \frac{i}{\hbar}\left\langle \hat{P}_{i}\frac{p_{k}\left(y_{k}\right)}{\left|\hat{x}_{i}-y_{k}\right|^{2}}+\frac{p_{k}\left(y_{k}\right)}{\left|\hat{x}_{i}-y_{k}\right|^{2}}\hat{P}_{i}\right\rangle \\
 & = & \int d\bm{x}\frac{p_{k}\left(y_{k}\right)}{\left|x_{i}-y_{k}\right|}\partial_{x_{i}}\left(\Psi\left(\bm{x}\right)^{*}\partial_{x_{i}}\Psi\left(\bm{x}\right)-\Psi\left(\bm{x}\right)\partial_{x_{i}}\Psi\left(\bm{x}\right)^{*}\right).
\end{eqnarray*}

This result can be connected to the continuity equation (which is satisfied by the SN theory):
\begin{equation}
\partial_{t}\rho+\nabla.\vec{j}=0,
\end{equation}
where
\begin{equation}
\rho=\left|\Psi\right|^{2};\qquad\vec{j}=\frac{\hbar}{2im}\left(\Psi^{*}\nabla\Psi-\Psi\nabla\Psi^{*}\right).
\end{equation}
We integrate over all variables except $x_{i}$ (which we denote by
$\bm{x}_{\ne i}$), obtaining 
\begin{eqnarray*}
\int d\bm{x}_{\ne i}\left(\partial_{t}\rho+\nabla.\vec{j}\right) & = & 0\\
 & = & \partial_{t}p_{i}\left(x_{i}\right)+\frac{\hbar}{2im}\int d\bm{x}_{\ne i}\sum_{j}\partial_{x_{j}}\left(\Psi^{*}\partial_{x_{j}}\Psi-\Psi\partial_{x_{j}}\Psi^{*}\right)
\end{eqnarray*}
For $j\ne i$, 
\begin{equation}
\int dx_{j}\partial_{x_{j}}\left(\Psi^{*}\partial_{x_{j}}\Psi-\Psi\partial_{x_{j}}\Psi^{*}\right)=0
\end{equation}
by integration by parts. Thus, 
\begin{equation}
\partial_{t}p_{i}\left(x_{i}\right)=-\frac{\hbar}{2im}\int d\bm{x}_{\ne i}\partial_{x_{i}}\left(\Psi^{*}\partial_{x_{i}}\Psi-\Psi\partial_{x_{i}}\Psi^{*}\right)
\end{equation}
so 
\begin{eqnarray*}
 &  & \frac{i}{\hbar}\left\langle \hat{P}_{i}\frac{p_{k}\left(y_{k}\right)}{\left|\hat{x}_{i}-y_{k}\right|^{2}}+\frac{p_{k}\left(y_{k}\right)}{\left|\hat{x}_{i}-y_{k}\right|^{2}}\hat{P}_{i}\right\rangle \\
 & = & \int dx_{i}\frac{p_{k}\left(y_{k}\right)}{\left|x_{i}-y_{k}\right|}\frac{-2im}{\hbar}\int d\bm{x}_{\ne i}\left(-\frac{\hbar}{2im}\partial_{x_{i}}\left(\Psi{}^{*}\partial_{x_{i}}\Psi-\Psi\partial_{x_{i}}\Psi{}^{*}\right)\right)\\
 & = & \frac{-2im}{\hbar}\int dx_{i}\frac{p_{k}\left(y_{k}\right)\dot{p}_{i}\left(x_{i}\right)}{\left|x_{i}-y_{k}\right|}.
\end{eqnarray*}
Substituting back into $d\left\langle \hat{E}\right\rangle /dt$,
\begin{eqnarray*}
\frac{d\left\langle \hat{E}\right\rangle }{dt} & = & \left(1-\beta\right)Gm^{2}\sum_{ji}\int dx_{i}\int dx_{j}\frac{p_{i}\left(x_{i}\right)\dot{p}_{j}\left(x_{j}\right)}{\left|x_{j}-x_{i}\right|}-\beta Gm^{2}\sum_{ij=1}^{N}\int dx_{i}\int dx_{j}\frac{p_{i}\left(x_{i}\right)\dot{p}_{j}\left(x_{j}\right)}{\left|x_{i}-x_{j}\right|},
\end{eqnarray*}
which is equal to 0 when 
\begin{equation}
1-\beta=\beta
\end{equation}
or $\beta=1/2$.

\section{\label{sec:spectraDerivation} Derivation of $p_0\rightarrow \xi$ and $p_0\leftarrow \xi$}

In this Appendix, we derive equations (\ref{eq:probPre}) and (\ref{eq:pxipost}) presented in subsection
\ref{sub:prePostNLQM}:  
\begin{eqnarray}
p_{0\rightarrow \xi} & \propto & \exp\left[-\frac{1}{2}\int\frac{d\Omega}{2\pi}\frac{|\xi(\Omega)-\langle \hat{b}_2(\Omega)\rangle_{0}|^2}{S_{A,A}}\right], \\
p_{0 \leftarrow \xi} & \propto & \exp\left[-\frac{1}{2}\int\frac{d\Omega}{2\pi}\frac{|\xi(\Omega)-\langle \hat{b}_2(\Omega)\rangle_{\xi}|^2}{S_{A,A}}\right]
.
\end{eqnarray}
They represent the probabilities of obtaining a particular measurement record 
\begin{equation}
\left\{ \xi\left(t\right):\;0 <t< \tau \right\} 
\end{equation}
over a period $\tau$ in the pre- and post-selection measurement prescriptions, respectively.

The probability of measuring $\xi(t)$ is
\begin{equation}
p_{\xi}=\left| \tensor[_{\rm out}]{\braOket{\xi}{\hat{U}}{0}}{_{\rm in}} \right|^2 \label{eq:probXi303},
\end{equation}
where $\hat{U}$ is a shorthand for the pre-selection time evolution operator
$\hat{U}_{\left|0\right\rangle _{{\rm in}}}$ or the post selection evolution operator $\hat{U}_{\left|\xi\right\rangle _{{\rm out}}}$, $\left|0\right\rangle _{{\rm in}}$ is a vacuum state for the incoming light, and $| \xi \rangle_{\rm out}$ is the state of the outgoing light corresponding to the measurement results $\xi(t)$.
We then rewrite $p_{\xi}$ to 
\begin{equation}
p_{\xi}=\left\langle 0|\hat{U}^{\dagger}|\xi\right\rangle \left\langle \xi|\hat{U}|0\right\rangle ,
\end{equation}
where we have used the shorthand $\left|\xi\right\rangle $ for $\left|\xi\right\rangle _{{\rm out}}$. $\hat{U}^{\dagger}\left|\xi\right\rangle \left\langle \xi\right|\hat{U}$
is a projection operator that can be written as a path integral (refer
to p.2 of \cite{khaliliNonGaussian} for a derivation): 
\begin{equation}
\hat{P}=\int\mathcal{D}k\left(t\right)\exp\left(i\int dtk\left(t\right)\left(\hat{b}_{2}\left(t\right)-\xi\left(t\right)\right)\right).
\end{equation}
Notice that in the limit that the SN non-linearity vanishes, $\hat{P}$ agrees with the standard quantum mechanics projector onto the measurement results $\xi(t)$. This is due to the fact that when $\osn$ vanishes, $\hat{b}_{2}$ becomes a linear operator which matches the prediction of standard quantum mechanics. Consequently, in the limit of $\osn\rightarrow 0$, $p_{0\rightarrow\xi}$ and $p_{0 \leftarrow \xi}$ recover the probabilities predicted by linear quantum mechanics.

Substituting $\hat{P}$ back into \eqref{probXi303}, we obtain
\begin{equation}
p_{\xi}=\int\mathcal{D}k\left(t\right)\braOket 0{\exp\left(i\int dtk\left(t\right)\hat{b}_{2}\left(t\right)\right)}0\exp\left(-i\int dtk\left(t\right)\xi\left(t\right)\right).
\end{equation}
Let us explicitly separate the mean of $\hat{b}_{2}\left(t\right)$ by defining $\hat{A}$ in the following way: $$\hat{b}_{2}\left(t\right)\equiv\hat{A}\left(t\right)+\braOket{0}{\hat{b}_{2}\left(t\right)}{0} \equiv \hat{A}\left(t\right)+\braket{\hat{b}_{2}\left(t\right)}.$$
We can then rewrite $p_\xi$ to
\begin{eqnarray}
p_{\xi} & = & \int\mathcal{D}k\left(t\right)\braOket 0{\exp\left(i\int dtk\left(t\right)\hat{A}\left(t\right)\right)}0\exp\left(-i\int dtk\left(t\right)\left(\xi\left(t\right)-\left\langle \hat{b}_{2}\left(t\right)\right\rangle \right)\right).
\end{eqnarray}
Next, we make use of the fact that $\left|0\right\rangle $ is a gaussian
state to rewrite the above expectation value as
\begin{equation}
p_{\xi}=\int\mathcal{D}k\left(t\right)\exp\left(-\frac{1}{2}\braOket 0{\left(\int dtk\left(t\right)\hat{A}\left(t\right)\right)^{2}}0\right)\exp\left(-i\int dtk\left(t\right)\left(\xi\left(t\right)-\left\langle \hat{b}_{2}\left(t\right)\right\rangle \right)\right)
\end{equation}
Expanding the first exponent, we obtain 
\begin{equation}
p_{\xi}=\int\mathcal{D}k\left(t\right)\exp\left(-\frac{1}{2}\int dt\int dzk\left(t\right)k\left(z\right)\left\langle \hat{A}\left(t\right)\hat{A}\left(z\right)\right\rangle \right)\exp\left(-i\int dtk\left(t\right)\left(\xi\left(t\right)-\left\langle \hat{b}_{2}\left(t\right)\right\rangle \right)\right).
\end{equation}
$p_{\xi}$ is a functional Gaussian integral over $k\left(t\right)$,
which we evaluate to 
\begin{equation}
p_{\xi}\propto\exp\left(-\frac{1}{2}\int dt\int dz
\left(\xi\left(t\right)-\left\langle \hat{b}_{2}\left(t\right)\right\rangle \right)
\left\langle \hat{A}\left(t\right)\hat{A}\left(z\right)\right\rangle^{-1}
\left(\xi\left(z\right)-\left\langle \hat{b}_{2}\left(z\right)\right\rangle \right)
\right), 
\end{equation}
where $\left\langle \hat{A}\left(t\right)\hat{A}\left(z\right)\right\rangle^{-1}$ is the inverse of the function $\left\langle \hat{A}\left(t\right)\hat{A}\left(z\right)\right\rangle$. 
Assuming we have a time-stationary process, $\left\langle \hat{A}\left(t\right)\hat{A}\left(z\right)\right\rangle $
can be simplified to $\left\langle \hat{A}\left(t-z\right)\hat{A}\left(0\right)\right\rangle $
which allows us to take a Fourier transform and obtain 
\begin{equation}
p_{\xi}\propto\exp\left(-\frac{1}{2}\int\frac{d\omega}{2\pi}\frac{\left|\xi\left(\omega\right)-\left\langle \hat{b}_{2}\left(\omega\right)\right\rangle \right|^{2}}{S_{A,A}\left(\omega\right)}\right).
\end{equation}
Finally, we note that for post-selection $\left\langle 0|\hat{b}_{2}\left(t\right)|0\right\rangle $ is calculated with $\hat{b}_{2}\left(t\right)$ obtained from an effective Heisenberg picture with the boundary state fixed to be the recorded eigenstates by the measurement device: $\left|\xi\right\rangle $.
For pre-selection, we obtain $\hat{b}_{2}\left(t\right)$ from an effective Heisenberg picture with the boundary state given to be the initial state of the light, vacuum.

\section{\label{sec:Appendix-Details-of-post-calculation} More details on calculating $\left\langle\hat{B}(\omega)\right\rangle_\xi$ }

In subsection \ref{sub:Post-Selection-analysis}, we calculated the spectrum of the outgoing light phase operator 
\begin{equation}
\hat{b}_{2}\left(\omega\right) 
=\hat{A}\left(\omega\right)+\left\langle \hat{B}\left(\omega\right)\right\rangle_{\xi},
\end{equation}
where we have neglected the contribution from classical thermal noise, as it is not important for this Appendix. 
Both $\hat A$ and $\hat B$ are linear operators of the form 
\begin{eqnarray}
\hat{A}\left(t\right) & = & \hat{a}_{2}\left(t\right)+\int_{-\infty}^{\infty}L_{A}\left(t-z\right)\hat{a}_{1}\left(z\right)dz\\
\hat{B}\left(t\right) & = & \int_{-\infty}^{\infty}L_{B}\left(t-z\right)\hat{a}_{1}\left(z\right)dz.
\end{eqnarray}
We presented their exact expressions in Eqs. (\ref{eq:Adef}) and (\ref{eq:Bdef}). Moreover, $\left\langle \hat{B}\left(\omega\right)\right\rangle_{\xi}$ is the expectation value of $\hat B$ over the outgoing light state $|\xi\rangle_{\rm out}$ corresponding to the measured eigenstates of the outgoing light's phase. In the calculation of the spectrum, and in particular of  $\left\langle \hat{B}\left(\omega\right)\right\rangle_{\xi}$, we stated without proof that if \eqref[name=Eq.~]{RAARzero} 
\begin{equation}
\tensor[_{\rm in}]{\braOket{0}{\hat{R}\left(t\right)\hat{A}\left(z\right)}{0}}{_{\rm in}} + 
\tensor[_{\rm in}]{\braOket{0}{\hat{A}\left(z\right)\hat{R}\left(t\right)}{0}}{_{\rm in}} = 0
\end{equation}
is satisfied then 
$$\tensor[_{\rm out}]{\braOket{\xi}{\hat{R}(t)}{\xi}}{_{\rm out}}\equiv\left\langle \hat{R}(t)\right\rangle _{\xi}=0$$
for all times $t$. $\hat{R}$ is defined by \eqref[name=Eq.~]{Bproj}. In this Appendix, we present the proof.

We first rewrite $\left|\xi\right\rangle _{{\rm out}}$ to
\begin{equation}
\left|\xi\right\rangle _{{\rm out}}=\hat{P}\left|0\right\rangle ,\label{eq:xiProj0}
\end{equation}
where 
\begin{equation}
\hat{P}\propto\int\mathcal{D}ke^{i\int dtk\left(t\right)\left(\hat{A}\left(t\right)-\eta\left(t\right)\right)}
\end{equation}
projects the initial state of the light, vacuum $\left|0\right\rangle $,
onto $\left|\xi\right\rangle _{{\rm out}}$. This form of $\hat{P}$
can be derived by referring to p.2 of \cite{khaliliNonGaussian} and
by making use of the fact that since $\left\langle \hat{B}\right\rangle _{\xi}$ is a $c$-number,
a measured eigenstate of $\hat{b}_{2}\left(t\right)$, $\left|\xi\left(t\right)\right\rangle $,
is also an eigenstate of $\hat{A}\left(t\right)$ with a different
eigenvalue which we choose to call $\eta\left(t\right)$.

Substituting \eqref[name=Eq.~]{xiProj0} into $\left\langle \hat{R}\left(t\right)\right\rangle _{\xi}$,
we obtain 
\begin{equation}
\left\langle \hat{R}\left(t\right)\right\rangle _{\xi}=\left\langle 0|\hat{P}\hat{R}\left(t\right)\hat{P}|0\right\rangle =\left.-i\partial_{\mu}\left\langle 0|\hat{P}e^{i\mu\hat{R}}\hat{P}|0\right\rangle \right|_{\mu=0}.
\end{equation}
Let us combine $P$ and $e^{i\mu\hat{R}}$  into one exponential by repeated
use of the Baker\textendash Campbell\textendash Hausdorff formula.
We begin with $\hat{P}e^{i\mu\hat{R}}$, 
\begin{eqnarray}
\hat{P}e^{i\mu\hat{R}} & = & \int\mathcal{D}ke^{i\int dtk\left(t\right)\left(\hat{A}\left(t\right)-\eta\left(t\right)\right)+i\mu\hat{R}}\exp\left(-\frac{\mu}{2}\int dzk\left(z\right)\left[\hat{A}\left(z\right),\hat{R}\left(t\right)\right]\right).
\end{eqnarray}
To evaluate the commutator, we make use of \eqref[name=Eq.~]{Bproj}
\begin{equation}
\hat{R}\left(t\right)=\hat{B}\left(t\right)-\int_{-\infty}^{T}K\left(t-z\right)\hat{A}\left(z\right)dz.
\end{equation}
Furthermore, since $A(t)$ and $B(t)$ are linear operators
\begin{eqnarray}
\left[\hat{A}\left(z\right),\hat{B}\left(t\right)\right] & = & \int_{-\infty}^{\infty}L_{B}\left(t-z\right)\left[\hat{a}_{2}\left(t\right),\hat{a}_{1}\left(z\right)\right] dz \\
 & = & -i\int_{-\infty}^{\infty}L_{B}\left(t-z\right)\delta\left(t-z\right) dz =-iL_{B}\left(0\right) .
\end{eqnarray}
Substituting this result back into $\hat{P}e^{i\mu\hat{R}}$, we obtain
\begin{eqnarray}
\hat{P}e^{i\mu\hat{R}} & = & \int\mathcal{D}ke^{i\int dtk\left(t\right)\left(\hat{A}\left(t\right)-\eta\left(t\right)\right)+i\mu\hat{R}}\exp\left(-\frac{i\mu L_{B}\left(0\right)}{2}\int dzk\left(z\right)\right)
\end{eqnarray}
Returning to $\hat{P}e^{i\mu\hat{R}}\hat{P}$, we have 
\begin{eqnarray*}
\hat{P}e^{i\mu\hat{R}}\hat{P} & = & \int\mathcal{D}l\int\mathcal{D}ke^{i\int dtk\left(t\right)\left(\hat{A}\left(t\right)-\eta\left(t\right)\right)+i\mu\hat{R}}e^{i\int dzl\left(z\right)\left(\hat{A}\left(z\right)-\eta\left(z\right)\right)}\exp\left(-\frac{i\mu L_{B}\left(0\right)}{2}\int dzk\left(z\right)\right)\\
 & = & \int\mathcal{D}l\int\mathcal{D}ke^{i\int dzk\left(z\right)\left(\hat{A}\left(z\right)-\eta\left(z\right)\right)+i\int dzl\left(z\right)\left(\hat{A}\left(z\right)-\eta\left(z\right)\right)+i\mu\hat{R}}\nonumber \\
 &  & \times\exp\left(-\frac{i\mu L_{B}\left(0\right)}{2}\int dzk\left(z\right)\right)\exp\left(\frac{\mu}{2}\int dzl\left(z\right)\left[\hat{A}\left(z\right),\hat{R}\left(t\right)\right]\right)\\
 & = & \int\mathcal{D}k_{+}e^{i\int dzk_{+}\left(z\right)\left(\hat{A}\left(z\right)-\eta\left(z\right)\right)+i\mu\hat{R}}\int\mathcal{D}k_{-}\exp\left(-\frac{i\mu L_{B}\left(0\right)}{2}\int dzk_{-}\left(z\right)\right)
\end{eqnarray*}
where we applied the Baker-Campbell-Hausdorff formula in the second line,  and in the third line, we defined $k_{+}=k\left(z\right)+l\left(z\right)$, and $k_{-}=k\left(z\right)-l\left(z\right)$.

Now, 
\begin{eqnarray}
\int\mathcal{D}k_{-}\exp\left(-\frac{i\mu L_{B}\left(0\right)}{2}\int dzk_{-}\left(z\right)\right) & = &  \lim_{n\rightarrow\infty} \delta^n\left(\frac{\mu L_{B}\left(0\right)}{2}\right) \equiv \prod\delta\left(\frac{\mu L_{B}\left(0\right)}{2}\right)
\end{eqnarray}
so
\begin{eqnarray*}
\partial_{\mu}\hat{P}e^{i\mu\hat{R}}\hat{P} & = & \left(\partial_{\mu}\int\mathcal{D}k_{+}e^{i\int dzk_{+}\left(z\right)\left(\hat{A}\left(z\right)-\eta\left(z\right)\right)+i\mu\hat{R}}\right)\times\prod\delta\left(\frac{\mu L_{B}\left(0\right)}{2}\right)\\
 &  & +\lim_{n\rightarrow\infty}n\int\mathcal{D}k_{+}e^{i\int dzk_{+}\left(z\right)\left(\hat{A}\left(z\right)-\eta\left(z\right)\right)+i\mu\hat{R}}\times\delta^{'}\left(\frac{\mu L_{B}\left(0\right)}{2}\right)\delta^{n-1}\left(\frac{\mu L_{B}\left(0\right)}{2}\right).
\end{eqnarray*}
When $\mu$ is set to 0, the second term will vanish because $\delta^{'}\left(\mu L_{B}\left(0\right)/2\right)$
vanishes at $\mu=0$ (as can be easily determined by writing the dirac-delta
function as a zero mean Gaussian with a vanishing variance). Consequently,
we only need to study the first term. 

Let take the expectation of $\partial_{\mu}\hat{P}e^{i\mu\hat{R}}\hat{P}$ over vacuum, 
\begin{eqnarray*}
\partial_{\mu}\braOket 0{\hat{P}e^{i\mu\hat{R}}\hat{P}}0  & = & \partial_{\mu}\int\mathcal{D}k_{+}\braOket 0{e^{i\int dzk_{+}\left(z\right)\hat{A}\left(z\right)+i\mu\hat{R}}}0 e^{-i\int dzk_{+}\left(z\right)\eta\left(z\right)}\times\prod\delta\left(\frac{\mu L_{B}\left(0\right)}{2}\right).
\end{eqnarray*}
We now analyze the first term in the integrand. Since $\left|0\right\rangle $ is a Gaussian state, the expectation over $\left|0\right\rangle $ can be simplified to 
\begin{eqnarray}
\braOket 0{e^{i\int dzk_{+}\left(z\right)\hat{A}\left(z\right)+i\mu\hat{R}}}0 & = & \exp\left(-\frac{1}{2}\mu^{2}\left\langle R^{2}\right\rangle \right)\exp\left(-\frac{1}{2}\mu\left\langle \hat{R}\times\int dzk_{+}\left(z\right)\hat{A}\left(z\right)\right\rangle +\left\langle \int dzk_{+}\left(z\right)\hat{A}\left(z\right)\times\hat{R}\right\rangle \right)\nonumber \\
 &  & \times\exp\left(-\frac{1}{2}\left\langle \left(dzk_{+}\left(z\right)\hat{A}\left(z\right)\right)^{2}\right\rangle \right).
\end{eqnarray}
The second exponential is equal to unity by the assumption given by \eqref[name=Eq.~]{RAARzero}. Thus, 
\begin{eqnarray*}
\braOket 0{e^{i\int dzk_{+}\left(z\right)\hat{A}\left(z\right)+i\mu\hat{R}}}0 & = & \exp\left(-\frac{1}{2}\mu^{2}\left\langle R^{2}\right\rangle \right)\exp\left(-\frac{1}{2}\left\langle \left(dzk_{+}\left(z\right)\hat{A}\left(z\right)\right)^{2}\right\rangle \right).
\end{eqnarray*}
Once we differentiate over $\mu$ and then set it to 0, this product
vanishes, giving 
\begin{equation}
\left\langle 0|\hat{R}|0\right\rangle =\left.\partial_{\mu}\braOket 0{\hat{P}e^{i\mu\hat{R}}\hat{P}}0\right|_{0}=0
\end{equation}
as desired.

\bibliography{references}

\end{document}